\documentclass[preprint,aps,amssymb,nofootinbib]{revtex4}
\usepackage{graphicx}
\usepackage{amsmath}
\usepackage{maybemath}
\usepackage{mathrsfs}
\usepackage{bm}
\usepackage[usenames,dvipsnames]{color}
\usepackage[colorlinks=true, pdfstartview=FitV, linkcolor=blue, citecolor=Orange,urlcolor=blue]{hyperref}

\def\diag{\mathop{\rm diag}\nolimits}
\def\Tr{\mathop{\rm Tr}\nolimits}
\def\sgn{\mathop{\rm sgn}\nolimits}
\def\Res{\mathop{\rm Res}\nolimits}
\def\Disc{\mathop{\rm Disc}\nolimits}
\def\Im{\mathop{\rm Im}\nolimits}

\def\PV{\mathop{\rm PV}\nolimits}
\def\partialsl{\mathop{\hspace{-0.1em}\partial\hspace{-0.55em}/\hspace{-0.05em}}\nolimits}
\def\Ksl{\mathop{K\hspace{-0.7em}/\hspace{0.2em}}\nolimits}
\def\Qsl{\mathop{Q\hspace{-0.7em}/\hspace{0.2em}}\nolimits}

\def\psl{\mathop{p\hspace{-0.45em}/}\nolimits}
\def\qsl{\mathop{q\hspace{-0.5em}/}\nolimits}
\def\ksl{\mathop{k\hspace{-0.5em}/}\nolimits}
\def\lsl{\mathop{l\hspace{-0.4em}/}\nolimits}

\begin{document}

\title{Bulk viscosity of the massive Gross-Neveu model}

\author{Daniel Fernandez-Fraile\footnote{\texttt{danfer@th.physik.uni-frankfurt.de}}}

\affiliation{Institut f\"ur Theoretische Physik, Johann Wolfgang Goethe-Universit\"at, Max-von-Laue-Str. 1, 60438 Frankfurt am Main, Germany.}

\date{September 14th, 2010}

\begin{abstract}
A calculation of the bulk viscosity for the massive Gross-Neveu model at zero fermion chemical potential is presented in the large-$N$ limit. This model resembles QCD in many important aspects: it is asymptotically free, has a dynamically generated mass gap, and for zero bare fermion mass it is scale invariant at the classical level (broken through the trace anomaly at the quantum level). For our purposes, the introduction of a bare fermion mass is necessary to break the integrability of the model, and thus to be able to study momentum transport. The main motivation is, by decreasing the bare mass, to analyze whether there is a correlation between the maximum in the trace anomaly and a possible maximum in the bulk viscosity, as recently conjectured. After numerical analysis, I find that there is no direct correlation between these two quantities: the bulk viscosity of the model is a monotonously decreasing function of the temperature. I also comment on the sum rule for the spectral density in the bulk channel, as well as on implications of this analysis for other systems.
\end{abstract}

\maketitle

\section{Introduction}

Transport coefficients are essential inputs to describe the space-time evolution of systems not far from
equilibrium. During the last years there has been a very
active effort to analyze them from both the theoretical and phenomenological points of
view in the context of heavy-ion collisions, condensed matter physics,
astrophysics and cosmology. The calculation of transport coefficients in quantum field theory at
intermediate and strong coupling is still a challenge from both the
analytical and the numerical points of view. Due to their intrinsic non-perturbative
nature, even in weakly interacting theories a resummation of an
infinite number of diagrams is needed in order to obtain the
leading-order result. In the
strongly coupled regime, the most prominent method available is the
AdS/CFT correspondence, although it is only applicable to a limited
class of field theories. On the other hand, lattice simulations are still not accurate enough regarding the calculation of spectral densities, and the introduction of a finite quark chemical potential makes things even more difficult because of the sign problem.

It was recently conjectured, based on a sum rule for the spectral density of the trace of the energy-momentum tensor in Yang-Mills theory \cite{Kha08}, that a maximum of the trace anomaly near the critical temperature might drive a maximum for the bulk viscosity near that temperature. The corresponding sum rule was later corrected in \cite{Rom09}, and the ansatz for the spectral density used to extract the bulk viscosity questioned \cite{Moo08,Car09,Rom09}. Since the trace anomaly measures the breaking of scale invariance in a system, and the bulk viscosity $\zeta$ essentially represents the difficulty for a system to relax back to equilibrium after a scale transformation, it seems in principle reasonable to think that $\zeta$ would be maximum when the breaking of scale invariance is maximum.

In heavy-ion phenomenology, bulk viscosity has usually been neglected because it is expected to be much smaller than the shear viscosity even at temperatures not very high \cite{Arn06a}. However, as suggested by the analysis of \cite{Kha08}, non-perturbative phenomena responsible for the main contribution to the trace anomaly near $T_\mathrm{c}$ could also produce a significant increase in the bulk viscosity. In this paper I will present an explicit calculation in the massive Gross-Neveu model in $1+1$ dimensions, where the correlation between trace anomaly and bulk viscosity can be accurately tested. I will not try to give an estimation for the absolute value of $\zeta$ in QCD near the phase transition though; as we will see this model is not suitable for that purpose. There are several works analyzing this issue employing different approaches (see for instance \cite{Pra93,Arn06a,Kar08,Gub08a,Che09,Fer09b,Sas09,Mey10} and references therein), but still the order of magnitude of the bulk viscosity near the crossover temperature is uncertain.

\vspace{0.3cm}In 1+1 dimensions, transverse flow of momentum is not possible, and the bulk viscosity is the only viscous coefficient present to linear order in gradients. In this paper I will analyze only finite temperature effects, considering a vanishing fermion chemical potential, thus the thermal conductivity will be zero in this case. Therefore, the only constitutive equation relevant for us is\footnote{I use the metric $g=\diag(+1,-1)$.}
\begin{align}\label{Eq:const11}
\langle\hat{T}^{11}\rangle=P_\mathrm{eq}-\zeta\,\frac{\partial u^1}{\partial x}\ ,
\end{align}
with $\langle\hat{T}^{11}\rangle$ the non-equilibrium expectation value for the spatial component of the energy-momentum tensor, $P_\mathrm{eq}$ is the pressure in equilibrium, and $u^1$ is the fluid velocity. The bulk viscosity can be in principle calculated perturbatively in field theory \cite{Jeo95}:
\begin{align}
\zeta\propto\lim_{\omega\rightarrow 0^+}\frac{\rho_\mathrm{bulk}(\omega)}{\omega}\ ,
\end{align}
where $\rho_\mathrm{bulk}$ is the spectral density corresponding to the thermal propagator $\langle T^\mu_\mu(t,x)T^\nu_\nu(0)\rangle$. Here though, I will use a kinetic theory approach, which should be equivalent to the diagrammatic one in the perturbative (and dilute) regime \cite{Jeo96}.

\vspace{0.3cm}This paper is organized as follows. In Sections \ref{Sec:MGNvac} and \ref{Sec:MGNtemp}, I review well known properties of the massive Gross-Neveu model at zero and finite temperature, and I prove the breaking of integrability in the large-$N$ limit when a mass term for the fermion field is explicitly introduced. Then in Section \ref{Sec:kinetictheory}, the calculation of the bulk viscosity within kinetic theory is presented. In Section \ref{Sec:discussion}, I comment on sum rules and implications of the previous analysis for other systems. Finally in \ref{Sec:conclusions} I summarize the main conclusions. There is also the Appendix \ref{App:fermions}, where the result of factorization for fermion loops in $1+1$ dimensions is derived, and Appendix \ref{App:amplitudes} where the reader can find some details on the calculation of the inelastic scattering amplitude.

\section{Vacuum properties of the massive Gross-Neveu model}\label{Sec:MGNvac}
Let's consider the Gross-Neveu model \cite{Gro74} with an explicit bare mass for the fermion field:
\begin{equation}
\mathcal{L}=\sum\limits_{a=1}^N\bar{\psi}_a\mathrm{i}\partialsl\psi_a+\frac{g^2}{2}\,\left(\sum\limits_{a=1}^N\bar{\psi}_a\psi_a-N m\right)^2\ .
\end{equation}
Since we are interested in studying the large $N$ limit of the model, in order for the perturbative expansion in powers of $1/N$ to be sensible, the bare coupling constant must be re-scaled, $g^2\equiv \lambda/N$, with $\lambda$ being constant as $N\rightarrow\infty$. Also, it is convenient to introduce an auxiliary field $\sigma$ to properly classify the different Feynman diagrams according to their topologies and power counting in $1/N$ \cite{Col74}:
\begin{equation}
\mathcal{L}=\sum\limits_{a=1}^N\bar{\psi}_a\mathrm{i}\partialsl\psi_a-\frac{1}{2}\,\sigma^2-g\sigma\sum\limits_{a=1}^N\bar{\psi}_a\psi_a+ N m g\sigma\ .
\end{equation}
Clearly, the introduction of this field does not affect the dynamics of the system because its equation of motion is simply $\sigma=N g m-g\sum_a\bar{\psi}_a\psi_a$. In terms of the auxiliary field, the discrete chiral symmetry then corresponds to the simultaneous transformations $\psi\mapsto\gamma_5\psi$ and $\sigma\mapsto-\sigma$.

In $1+1$ space-time dimensions and in the large-$N$ limit, this model shares many important features with massless QCD in $3+1$ dimensions: it is renormalizable, asymptotically free, classically scale invariant (for zero bare fermion mass), it has a dynamically generated mass gap which manifests as a peak in the trace anomaly, and in vacuum undergoes an spontaneous breaking of the discrete ``chiral'' symmetry\footnote{In massless QCD instead, it is the continuous chiral symmetry $\mathrm{SU}(N_\mathrm{f})_\mathrm{A}\times \mathrm{SU}(N_\mathrm{f})_\mathrm{V}$ which is spontaneously broken in vacuum down to $\mathrm{SU}(N_\mathrm{f})_\mathrm{V}$.} $\psi\mapsto\gamma_5\psi$.

As we will see in the next subsection, the introduction of this bare mass $m$ is a simple way of allowing the system to relax back to thermodynamic equilibrium after a small perturbation in the distribution of momenta. In addition, the bare mass also suppresses the density of kink-anti-kink configurations in the thermodynamic limit and makes the mean-field $1/N$ expansion well defined \cite{Das75a,Bar95}.

\begin{figure}[h!]
\begin{center}
\includegraphics[width=8cm]{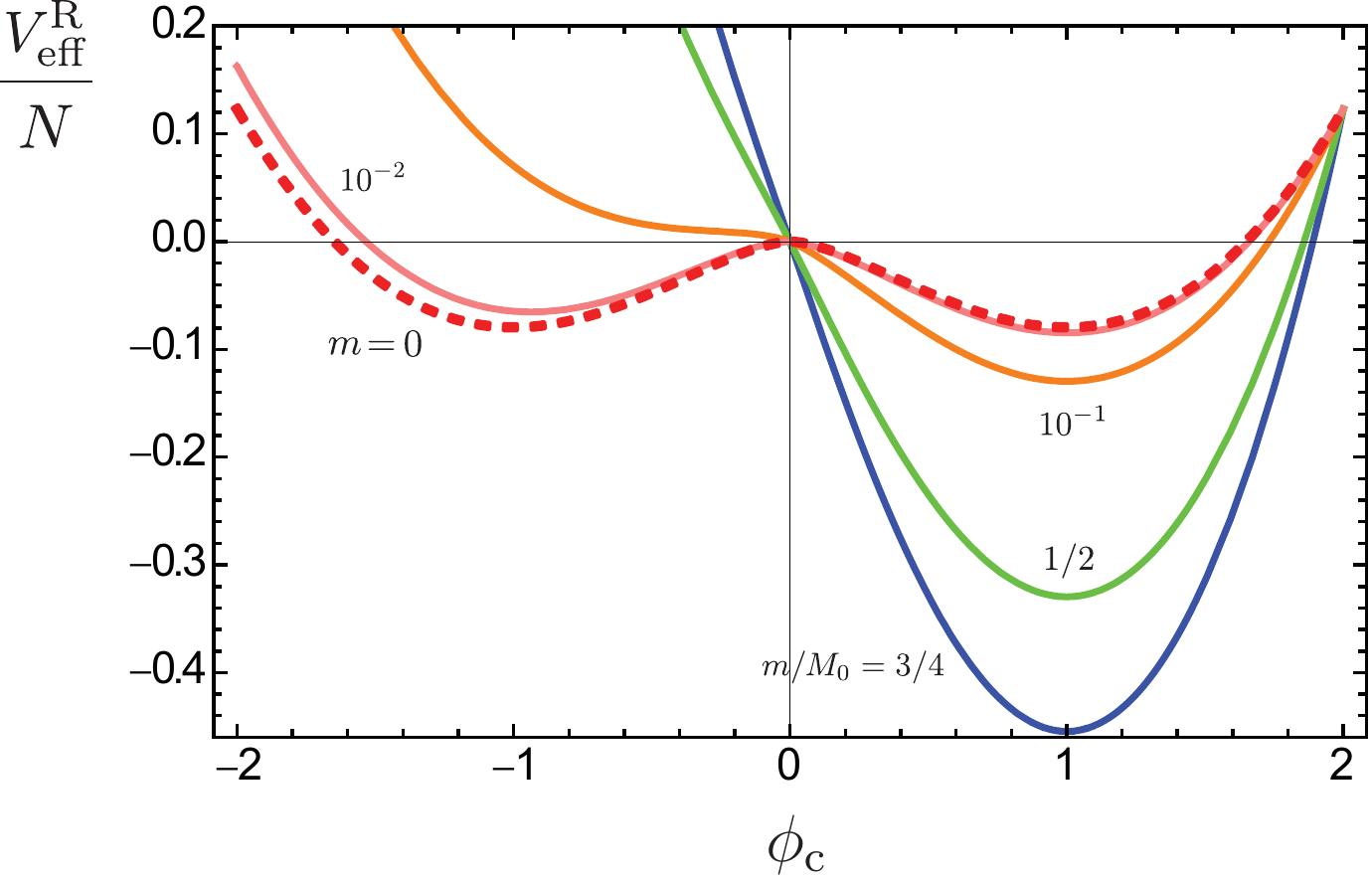}
\caption{Effective potential of the classical field $\phi_\mathrm{c}$ for different values of $m$.}\label{Fig:potential}
\end{center}
\end{figure}

To leading order in the large-$N$ limit, only one counter-term is necessary to renormalize all the divergences, $\delta\mathcal{L}=\delta m_\sigma\sigma^2/2$, which essentially amounts to a renormalization of the coupling constant. The effective potential for the classical field $\sigma_\mathrm{c}$ is obtained using standard techniques and renormalized imposing the condition $\mathrm{d}^2V_\mathrm{eff}(\sigma_\mathrm{c})/\mathrm{d}\sigma_\mathrm{c}^2|_{\sigma_\mathrm{c}=\sigma_0}=1$ \cite{Gro74}, with $\sigma_0$ the renormalization scale. This fixes the counter-term to be
\begin{equation}
\delta m_\sigma^2 = \frac{g^2N}{2\pi}\left[\ln\left(\frac{\mathnormal{\Lambda}}{g\sigma_0}\right)^2 -2\right]\ ,
\end{equation}
with $\mathnormal{\Lambda}$ an ultraviolet cutoff. Then, the leading-order renormalized effective potential is
\begin{equation}\label{Eq:renpot}
V_\mathrm{eff}^\mathrm{R}(\sigma_\mathrm{c})=\frac{1}{2}\,\sigma_\mathrm{c}^2- N m g \sigma_\mathrm{c}+\frac{g^2 N\sigma_\mathrm{c}^2}{4\pi} \left[\ln\left(\frac{\sigma_\mathrm{c}}{\sigma_0}\right)^2-3\right]\ .
\end{equation}
This is a Mexican-hat potential (tilted by the mass $m$) with a non-zero mass gap $M_0$ determined by the condition
\begin{equation}\label{Eq:gapeq0}
\left.\frac{\mathrm{d}V_\mathrm{eff}^\mathrm{R}(\sigma_\mathrm{c})}{\mathrm{d}\sigma_\mathrm{c}}\right|_{\sigma_\mathrm{c} =M_0/g}=0\quad\Rightarrow\quad M_0-g^2 N m+\frac{g^2 N M_0}{2\pi}\left[\ln\left(\frac{M_0}{g\sigma_0}\right)^2-2\right]=0\ .
\end{equation}
If we define $\phi_\mathrm{c}\equiv g\sigma_\mathrm{c}$, we can now use (\ref{Eq:gapeq0}) to write the effective potential in a scale-independent form:
\begin{align}\label{Eq:effpot}
V_\mathrm{eff}^\mathrm{R}(\phi_\mathrm{c})=N m \phi_\mathrm{c}\left(\frac{\phi_\mathrm{c}}{2 M_0}-1\right)+\frac{N \phi_\mathrm{c}^2}{4\pi}\left[\ln\left(\frac{\phi_\mathrm{c}}{M_0}\right)^2-1\right]\ .
\end{align}
As shown in Fig. \ref{Fig:potential}, for $m=0$ the discrete chiral symmetry is spontaneously broken by choosing as vacuum one of the two minima. For small enough values of $m$, the potential still has two minima, whereas for larger $m$ one disappears and the other one becomes deeper (I use the units $M_0\equiv 1$).

The effective potential (\ref{Eq:renpot}) satisfies the renormalization-group equation
\begin{align}
\left[\sigma_0\frac{\partial}{\partial\sigma_0}+\beta(g)\frac{\partial}{\partial g}-\gamma_\sigma(g)\sigma_\mathrm{c}\frac{\partial}{\partial\sigma_\mathrm{c}}\right]V_\mathrm{eff}^\mathrm{R}(\sigma_\mathrm{c},g,\sigma_0)=0\ ,
\end{align}
which implies
\begin{align}
\beta(g)=g\gamma_\sigma(g)=-\frac{g^3N/2\pi}{1+g^2N/2\pi}\ ,
\end{align}
i.e., the theory is asymptotically free. Although the running coupling constant becomes arbitrarily large at low energies, the interaction between fermions is also suppressed by powers of $1/N$, thus in the large-$N$ limit we are still able to probe the low-energy regime of the theory.

\begin{figure}[h!]
\begin{center}
\includegraphics[width=6cm]{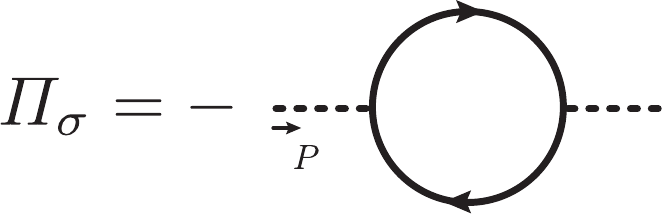}
\caption{Self-energy of the $\sigma$ field to leading order, $\mathcal{O}(N^0)$.}\label{Fig:self-energy}
\end{center}
\end{figure}

The leading-order contribution to the self-energy of the $\sigma$ field corresponds to the diagram depicted in Fig. \ref{Fig:self-energy}. In Euclidean space, the expression for the (renormalized) $\sigma$ propagator in vacuum is
\begin{align}\label{Eq:sigmapvacuum}
[D_{\sigma,\mathrm{E}}^0(P)]^{-1}=g^2N\frac{m}{M_0}+\frac{g^2N}{2\pi}\beta(P^2)\ln\left[\frac{\beta(P^2)+1}{\beta(P^2)-1}\right]\ ,
\end{align}
with $\beta(P^2)\equiv\sqrt{1+4M_0^2/P^2}$ a phase-space factor, and $P\equiv(p_1,p_2)$.

In the next subsection, I show how the first term in (\ref{Eq:sigmapvacuum}) breaks the integrability of the model in the large-$N$ limit.

\subsection{Breaking of integrability}\label{Sec:integrability}
The Gross-Neveu model (without the bare mass) is an integrable quantum field theory \cite{Zam79,Wit78}, this implies the existence of an infinite number of conserved charges and $1+1$ dimensions the factorization of the $S$-matrix in terms of binary collisions, so inelastic processes have vanishing scattering amplitude. Since in $1+1$ dimensions binary collisions cannot modify the distribution of momenta, integrability then prevents momentum transport in this system. Consequently, the bulk viscosity of the Gross-Neveu model is infinite. After including the bare mass in the model, this factorization in terms of binary collisions no longer happens, and hence it renders the bulk viscosity finite.

\begin{figure}[h!]
\begin{center}
\includegraphics[width=12cm]{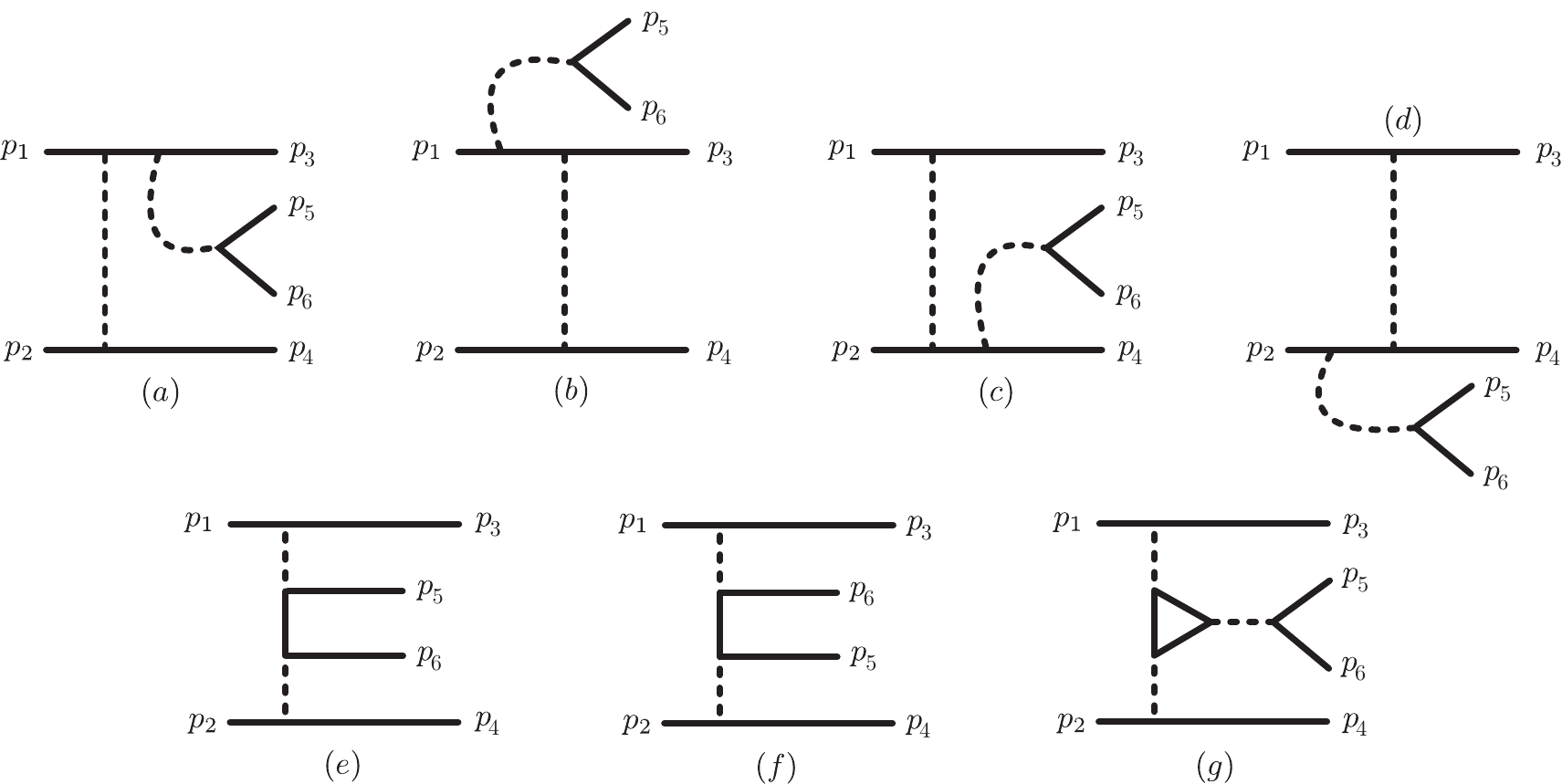}
\caption{Leading-order contribution in the large-$N$ limit to the inelastic process $1 2\rightarrow 3 4 5 6$.}\label{Fig:leadinginel}
\end{center}
\end{figure}

\begin{figure}[h!]
\begin{center}
\includegraphics[width=11cm]{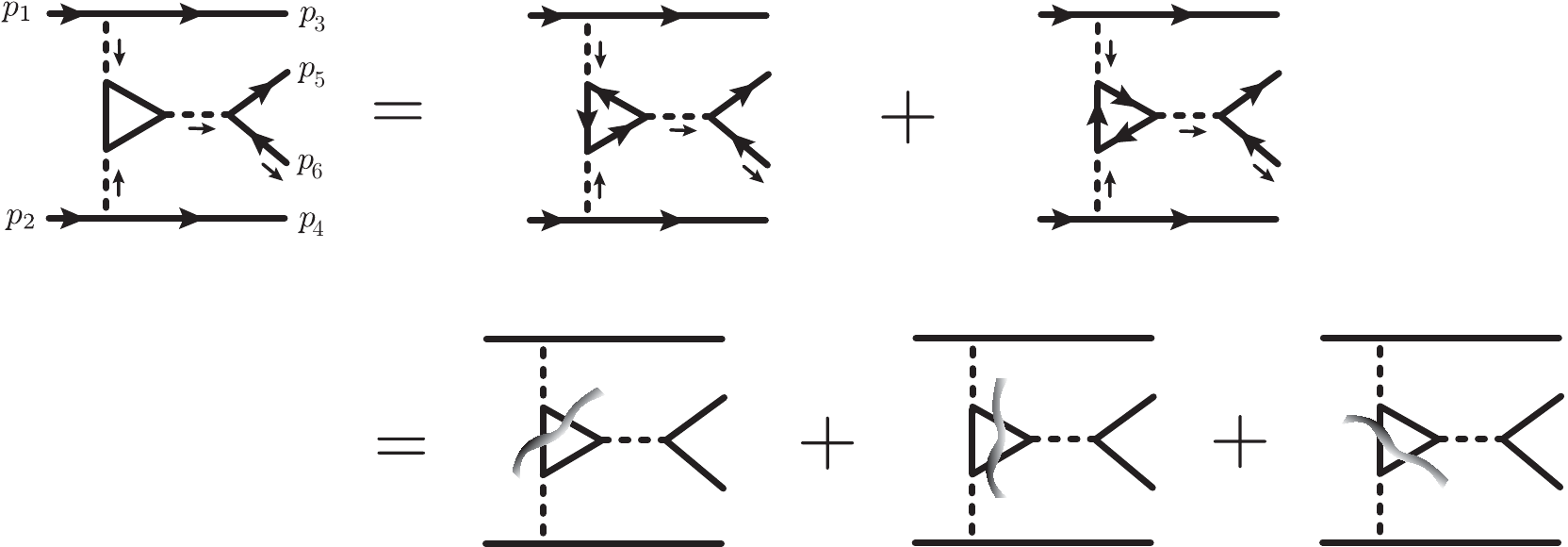}
\caption{Diagram of Fig. \ref{Fig:leadinginel}(g) expressed in terms of different cuts according to Eq. (\ref{Eq:fermioncut}).}\label{Fig:diaggcut}
\end{center}
\end{figure}

\begin{figure}[h!]
\begin{center}
\includegraphics[width=14cm]{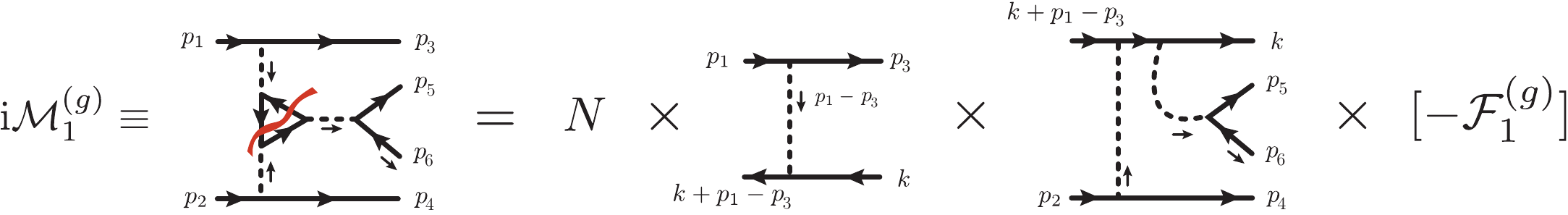}
\caption{Factorization in terms of tree diagrams.}\label{Fig:diaggcut1}
\end{center}
\end{figure}

To see this, consider the leading-order diagrams corresponding to the inelastic process $2\rightarrow 4$ in Fig. \ref{Fig:leadinginel}. As it was shown in \cite{Zam79}, the fermion loop of Fig. \ref{Fig:leadinginel}(g) factorizes into tree diagrams corresponding to all the possible ways of cutting it (Figs. \ref{Fig:diaggcut} and \ref{Fig:diaggcut1}). One particular cut is depicted in Fig. \ref{Fig:diaggcut1}. From the result (\ref{Eq:defF}) derived in Appendix \ref{App:fermions}, it is easy to see that the four-point amplitude and the factor $-\mathcal{F}$ in Fig. \ref{Fig:diaggcut1} cancel out giving a $-1$ factor. Hence, the diagram of Fig. \ref{Fig:diaggcut1} exactly cancels (when $m=0$) the one of Fig. \ref{Fig:leadinginel}(a), and the same for the rest of diagrams. If we now introduce the mass $m$, from (\ref{Eq:sigmapvacuum}) we see that this cancellation cannot happen, so the total inelastic amplitude is now $\propto m/M_0$ to leading order in $1/N$. This proves the non-integrability of the massive Gross-Neveu model in the large-$N$ limit.

%

\section{The model at finite temperature}\label{Sec:MGNtemp}

The thermodynamic properties of this model have been studied in detail in many papers, see for instance \cite{Kli86,Bar95,Sch00,Bla03,Sch06} and references therein. In this section, I am simply going to review leading order-results in the mean-field approximation, which are relevant for the later analysis of the bulk viscosity.

The leading-order renormalized effective potential at finite temperature is
\begin{equation}
V_\mathrm{eff}^\mathrm{R}(\sigma_\mathrm{c};T)=\frac{1}{2}\,\sigma_\mathrm{c}^2- g N m \sigma_\mathrm{c}+\frac{g^2 N\sigma_\mathrm{c}^2}{4\pi} \left[\ln\left(\frac{\sigma_\mathrm{c}}{\sigma_0}\right)^2-3\right]- \frac{2 N T}{\pi}\int\limits_0^\infty\mathrm{d}k\ \ln\left(1+\mathrm{e}^{-\sqrt{k^2+g^2 \sigma_\mathrm{c}^2}/T}\right)\ .
\end{equation}

\begin{figure}[h!]
\includegraphics[width=8cm]{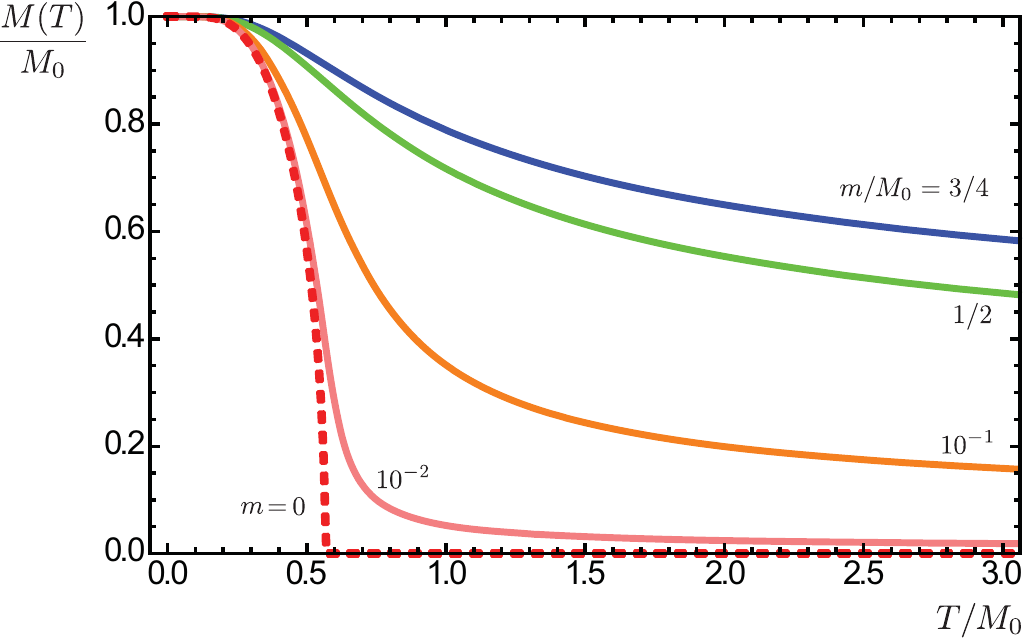}
\caption{Thermal mass gap of the fermion field as a function of the temperature for different values of the quotient $m/M_0$.}\label{Fig:thermalmass}
\end{figure}

The thermal mass gap is defined by
\begin{align}
&\left.\frac{\mathrm{d}V_\mathrm{eff}^\mathrm{R}(\sigma_\mathrm{c};T)} {\mathrm{d}\sigma_\mathrm{c}}\right|_{\sigma_\mathrm{c} =M(T)/g}=0\notag\\
&\Rightarrow\quad m\left(\frac{1}{M_0}-\frac{1}{M(T)}\right)+ \frac{1}{2\pi}\ln\left(\frac{M(T)}{M_0}\right)^2+\frac{2}{\pi} \int\limits_0^\infty\mathrm{d}k\ \frac{n_\mathrm{F}(E_k)}{E_k}=0\ ,
\end{align}
where (\ref{Eq:gapeq0}) has been used, $E_k\equiv \sqrt{k^2+M(T)^2}$, and $n_\mathrm{F}(x)\equiv(\exp(x/T)+1)^{-1}$ is the Fermi-Dirac distribution function. In Fig. \ref{Fig:thermalmass}, I plot the fermion mass gap as a function of the temperature for different values of $m$. For the case $m=0$, the mass gap vanishes at the temperature $T_\mathrm{c}\simeq 0.57 M_0$, indicating restoration of the discrete chiral symmetry. This is however an artifact of the mean-field approximation; the chiral symmetry is actually immediately restored at $T=0^+$ due to kink-anti-kink configurations\footnote{This restoration must happen in order to be consistent with the Mermin-Wagner theorem \cite{Mer66}. Note however that the phase transition at the indicated critical temperature occurs and is correctly reproduced in the mean-field approximation if the size of the system is kept finite and the limit $N\rightarrow\infty$ is taken first \cite{Sch00}.}. Nevertheless, as mentioned above, the introduction of a finite bare mass suppresses these kink-anti-kink configurations in the thermodynamic limit, and therefore we can approach in the mean-field approximation the curve $m=0$ as much as we wish provided we keep $m$ finite\footnote{Here it is important to emphasize that in our calculations the large-$N$ and thermodynamic limits are taken first keeping $m$ finite, and afterwards we study the limit $m\rightarrow 0^+$.}.

The pressure is immediately obtained from the effective potential:

\begin{align}
P=-V_\mathrm{eff}^\mathrm{R}(M(T)/g;T)=&\, m N M(T) \left(1-\frac{M(T)}{2 M_0}\right)-\frac{N M(T)^2}{4\pi}\left[\ln\left(\frac{M(T)}{M_0}\right)^2-1\right]\notag\\
&+\frac{2 N T}{\pi}\int\limits_0^\infty\mathrm{d}k\ \ln\left(1+\mathrm{e}^{-\sqrt{k^2+M(T)^2}/T}\right)\ ,
\end{align}
and the ``bag pressure'' is
\begin{equation}
P_\mathrm{bag}\equiv P(T=0)=\frac{N M_0^2}{2}\left(\frac{1}{2\pi}+\frac{m}{M_0}\right)>0\ .
\end{equation}

\begin{figure}[h!]
\includegraphics[width=8cm]{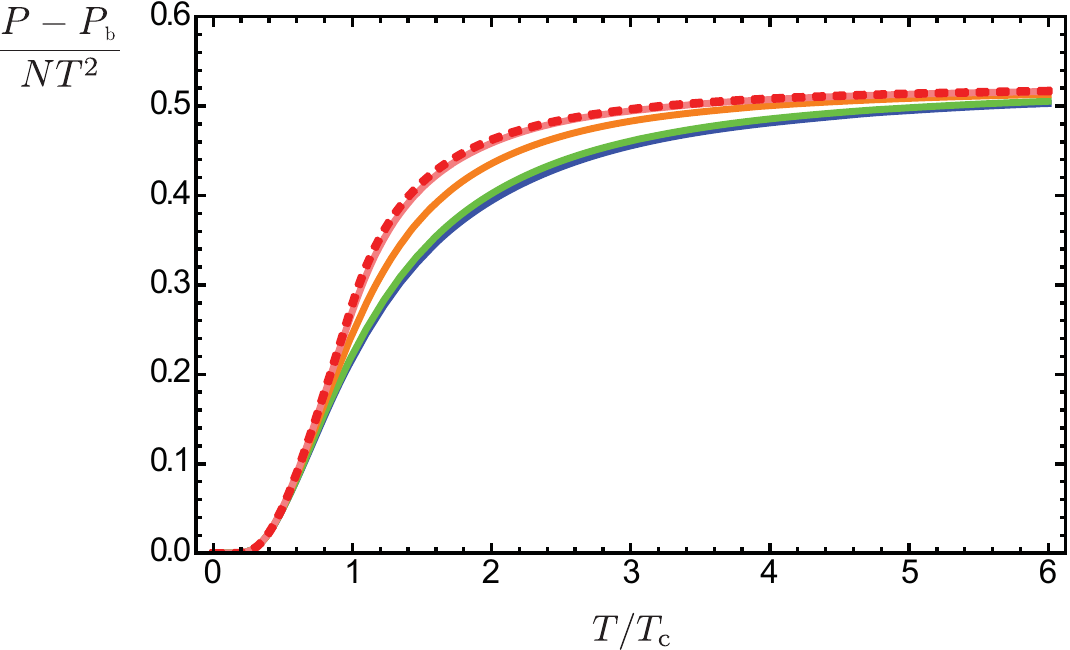}
\caption{Pressure as a function of the temperature for different values of $m/M_0$. The color code is the same as in Fig. \ref{Fig:thermalmass}.}\label{Fig:pressure}
\end{figure}

\begin{figure}[h!]
\includegraphics[width=8cm]{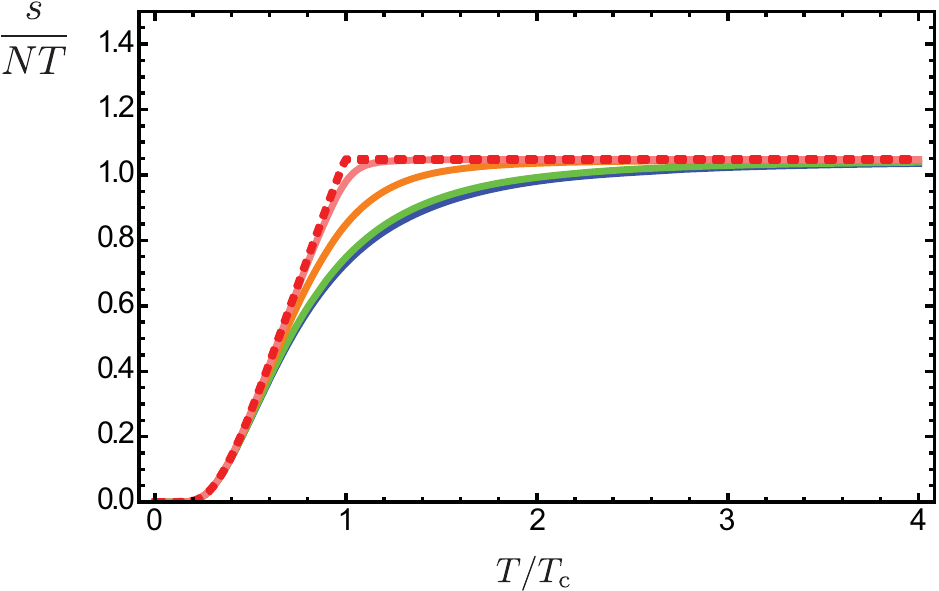}
\caption{Entropy density.}\label{Fig:entropy}
\end{figure}

\begin{figure}[h!]
\includegraphics[width=8cm]{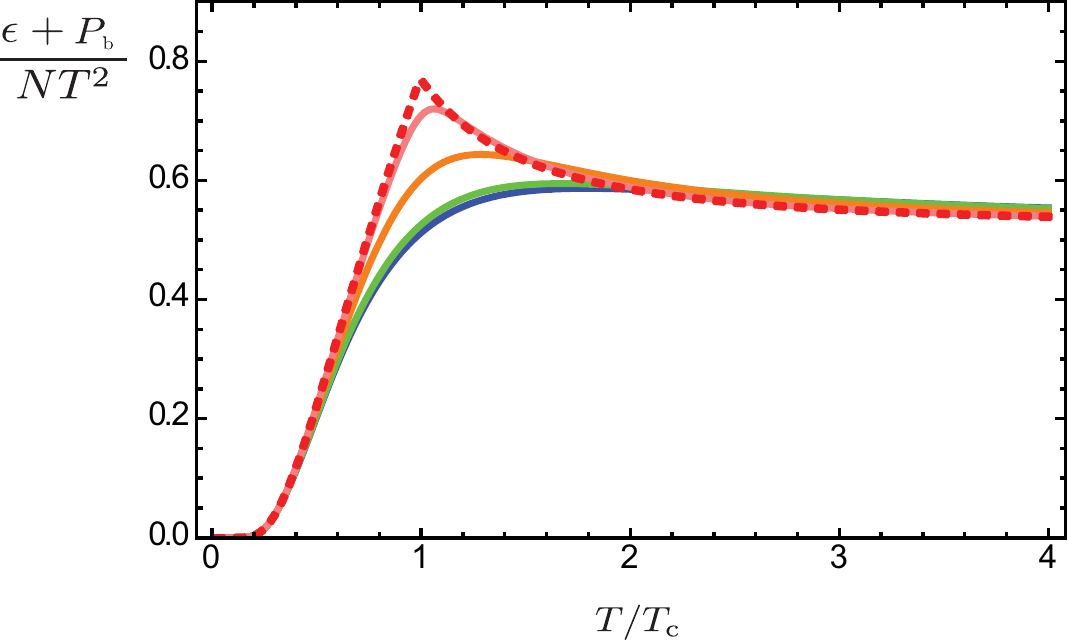}
\caption{Energy density.}\label{Fig:energydens}
\end{figure}

\begin{figure}[h!]
\includegraphics[width=8cm]{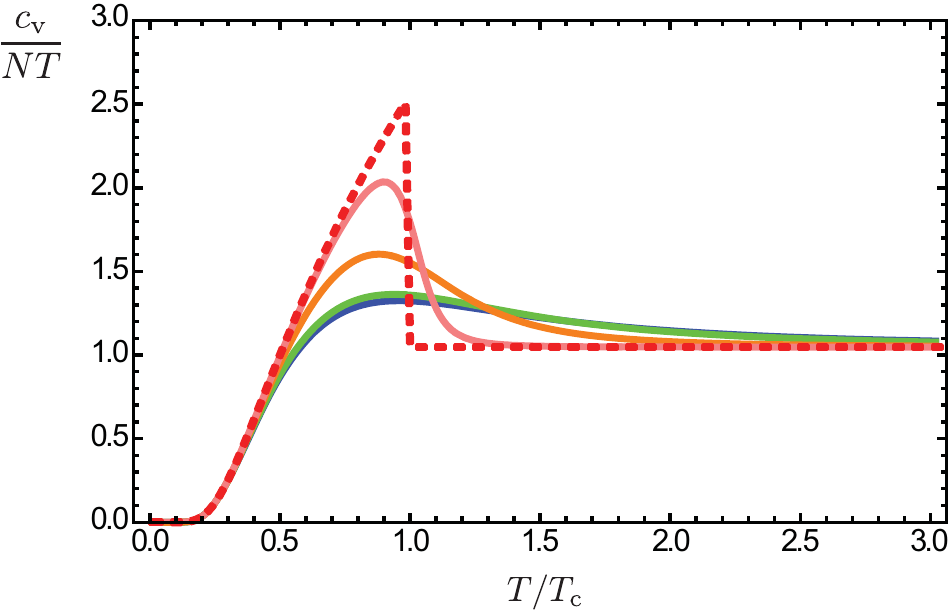}
\caption{Specific heat at constant volume. For $m=0$, $c_\mathrm{v}$ has a discontinuity at $T=T_\mathrm{c}$.}\label{Fig:specificheat}
\end{figure}

\begin{figure}[h!]
\includegraphics[width=8cm]{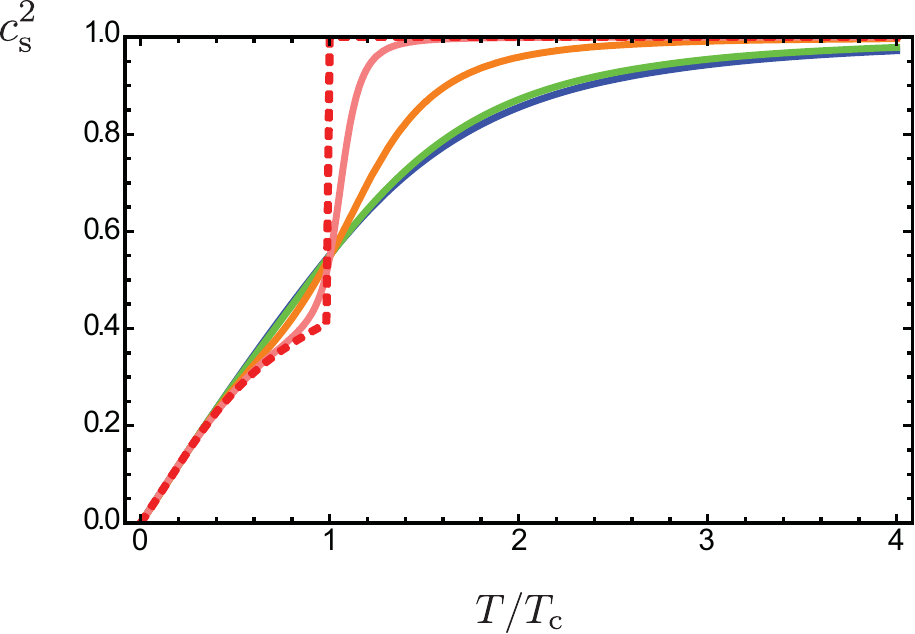}
\caption{Speed of sound squared. For $m=0$, $c_\mathrm{c}^2$ has a discontinuity at $T=T_\mathrm{c}$.}\label{Fig:speedsound}
\end{figure}

\begin{figure}[h!]
\includegraphics[width=8cm]{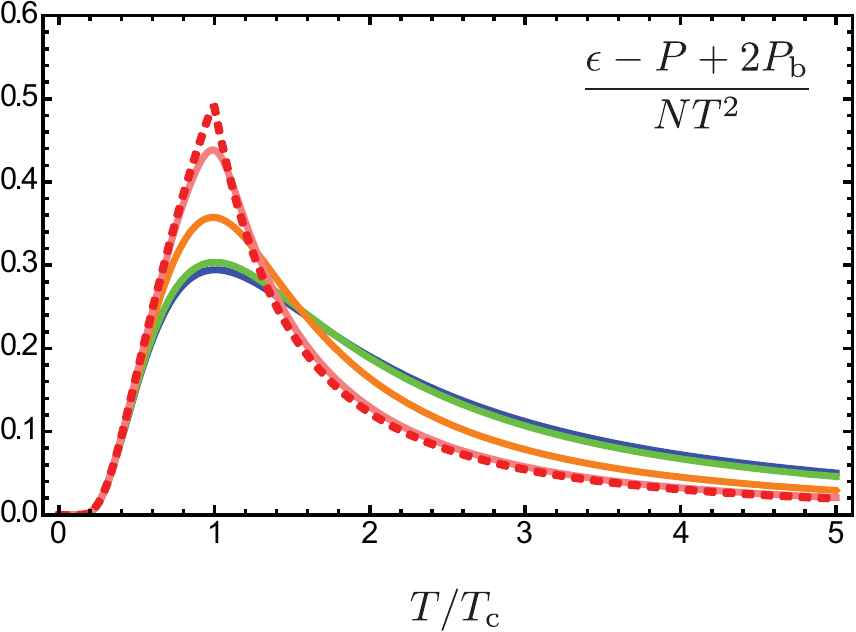}
\caption{Trace anomaly.}\label{Fig:traceanom}
\end{figure}

Entropy, energy density, specific heat, speed of sound, and trace anomaly are calculated from the pressure using the thermodynamic relations
\begin{align}
& s=\frac{\partial P}{\partial T}\ ,\quad \epsilon=Ts-P=T^2\frac{\partial}{\partial T}\left(\frac{P}{T}\right)\ ,\quad c_\mathrm{v}=\frac{\partial\epsilon}{\partial T}=T\frac{\partial s}{\partial T}\ ,\quad c_\mathrm{s}^2=\frac{\partial P}{\partial\epsilon}=\frac{s}{c_\mathrm{v}}\ ,\notag\\
& \mathnormal{\Delta}\equiv\frac{\epsilon-P+2P_\mathrm{b}}{T^2}=T\frac{\partial}{\partial T}\left(\frac{P-P_\mathrm{b}}{T^2}\right)\ .
\end{align}
These are plotted in Figs. \ref{Fig:pressure}, \ref{Fig:entropy}, \ref{Fig:energydens}, \ref{Fig:specificheat}, \ref{Fig:speedsound}, \ref{Fig:traceanom}. We see that the trace anomaly has a very pronounced peak right at $T_\mathrm{c}$ for $m=0^+$, which will allow us to study the possible correlation with the bulk viscosity. For $m=0^+$, above $T_\mathrm{c}$ the pressure corresponds to an ideal gas of massless fermions:
\begin{align}
P=\frac{\pi N T^2}{6}\ ,\quad \epsilon=P\ ,\quad s=\frac{\pi N T}{3}\ ,\quad c_\mathrm{v}=s\ , \quad c_\mathrm{s}^2=1\ ,\quad\mathnormal{\Delta}=\frac{2P_\mathrm{b}}{T^2}\ ,
\end{align}
with $P_\mathrm{b}=NM_0^2/(4\pi)$.


 In order to calculate dynamical quantities, it is convenient to shift $\sigma\mapsto M(T)/g+\sigma$, so tadpole diagrams vanish and have not to be taken into account. The $\sigma$ propagator to leading order and at finite temperature, calculated from the diagram of Fig. \ref{Fig:self-energy} in the Imaginary-Time Formalism and continued to real frequencies, is

\begin{align}\label{Eq:sigmapT}
&\frac{1}{g^2N}[\mathnormal{\Delta}_\sigma(\mathrm{i}\omega_n\mapsto\omega+\mathrm{i}0^+,p)]^{-1}=\frac{m}{M}+\frac{1}{2\pi}\left\{[\theta(-s)+\theta(s-4 M^2)]\beta(s)\ln\left|\frac{\beta(s)+1}{\beta(s)-1}\right|\right.\notag\\
&\left.+\theta(s)\theta(4M^2-s)2B(s)\arctan\left(\frac{1}{B(s)}\right)\right\}+\PV\int\limits_{-\infty}^\infty\frac{\mathrm{d}k}{\pi}\ \frac{n_\mathrm{F}(E_k)}{E_k}\frac{s\beta(s)^2(2pk-s)}{(2pk-s)^2-4E_k^2\omega^2} \notag\\
&-\frac{\mathrm{i}}{2}\sgn(\omega)\beta(s)\left\{\theta(s-4M^2)[1-n_\mathrm{F}(\epsilon_+)-n_\mathrm{F}(\epsilon_-)]+ \theta(-s)|n_\mathrm{F}(\epsilon_+)-n_\mathrm{F}(\epsilon_-)|\right\}\ ,
\end{align}
where $M\equiv M(T)$, $\beta(s)\equiv\sqrt{1-4M^2/s}$, $B(s)\equiv\sqrt{4M^2/s-1}$, and $\epsilon_\pm\equiv|\omega\pm p\beta(s)|/2$.

The first term in (\ref{Eq:sigmapT}) is responsible of breaking the integrability of the model also at finite temperature, which follows from the result (\ref{Eq:fermioncut}) in Appendix \ref{App:fermions} in the same way as for the vacuum case analyzed in the previous section. We realize that the breaking of integrability is now controlled by the factor $m/M(T)$, instead of $m/M_0$. Interestingly, for $T>T_\mathrm{c}$, the limit of $m/M(T)$ as $m\rightarrow 0^+$ is not zero, but a (temperature-dependent) constant. Thus, the scattering amplitude for inelastic processes is always finite when $m=0^+$ for $T>T_\mathrm{c}$.

\section{Kinetic theory approach}\label{Sec:kinetictheory}
The massive Gross-Neveu model is a non-confining theory and, as we have seen in the previous sections, the interaction between the fundamental fermions is suppressed by powers of $1/N$, hence in principle it seems reasonable to adopt a kinetic theory treatment to analyze the transport properties of this system in the large-$N$ limit. Alternatively, one could formally work out the resummation of an infinite series of ladder and chain diagrams contributing the spectral density of the energy-momentum tensor when the external frequency goes to zero. However, it is known that this resummation leads to solving an integral equation which coincides with the Boltzmann equation in the effective kinetic theory describing thermal excitations in the system \cite{Jeo95,Jeo96,Arn00,Aar04,Dob09}.

I consider that the kinetic theory approach is simpler, and I will employ it for the calculation of the bulk viscosity in this paper\footnote{Strictly, due to infrared divergencies characteristic of low-dimensional systems, this calculation is valid in $1+1$ dimensions only in the limit $N\rightarrow\infty$, where the long-time tail in the energy-momentum tensor correlator $\sim t^{-1/2}$ becomes negligible \cite{Kov03}. Otherwise, for $N$ finite, the bulk viscosity of the massive Gross-Neveu model would be infinite.}. I am going to follow essentially the previous works \cite{Arn06a,Arn00,Arn03,Jeo95,Jeo96} so, although I try to keep the discussion self-contained, the reader is referred to these papers for additional details.

In order to obtain the bulk viscosity, we need to determine the statistical average of the energy-momentum tensor of the system in a cell of fluid for a small departure from equilibrium. In kinetic theory, this average is \cite{Gro80}
\begin{align}\label{Eq:emtensor}
T^{\mu\nu}(t,x)=\sum\limits_A\int\limits_{-\infty}^\infty\frac{\mathrm{d}k}{(2\pi) E_k}\, \underline{k}^\mu \underline{k}^\nu f^A(t,x,\underline{k})\ ,
\end{align}
where $f^A= f^A(t,x,\underline{k})$ is the non-equilibrium distribution function, $A$ is a collective index denoting the fermionic or anti-fermionic character and the flavor of the corresponding particle species, $E_k\equiv\sqrt{M(T)^2+k^2}$, and $\underline{k}=(E_k,k)$ is the canonical momentum (the underline emphasizes that it is on-shell).

The Boltzmann-Uehling-Uhlenbeck equation determines the space-time evolution of distribution functions for dilute systems due to the change in the number of particles of type $A$ produced by collisions in the fluid. In 1+1 dimensions it reads
\begin{align}\label{Eq:Boltzmann}
\left(\frac{\partial}{\partial t}+\frac{k}{E_k}\frac{\partial}{\partial x}\right)f^A=\left.\frac{\partial f^A}{\partial t}\right|_\mathrm{gain}-\left.\frac{\partial f^A}{\partial t}\right|_\mathrm{loss}\equiv \frac{1}{E_k}\,\mathcal{C}^A_k[\bm{f}]\ ,
\end{align}
with $\bm{f}=(\{f^A\})$ a column vector containing the distribution functions for every type of particle. Considering only the leading-order elastic and inelastic processes in the large-$N$ expansion, which in our case are $1\,2\leftrightarrow 3\,4$ or $1\,2\,3\leftrightarrow 4\,5\,6$, and $1\,2\leftrightarrow 3\,4\,5\,6$ or $1\,2\,3\,4\leftrightarrow 5\,6$ respectively, the collision term is given by
\begin{align}
&\mathcal{C}^A_1[\bm{f}]\,\frac{\mathrm{d}p_1}{2\pi E_1}=\sum\limits_{B,C,D}\,\int\limits_{2,3,4}\frac{\mathrm{d}p_1}{2\pi}\frac{\mathrm{d}p_2}{2\pi}\, \mathrm{d}\mathnormal{\Gamma}_{1 2\rightarrow 3 4}^{A,B;C,D} L\, S_{(1) 2\rightarrow 3 4}^{A,B;C,D}\, [f^C_3 f^D_4 (1-f^A_1)(1-f^B_2)\notag\\
&-f^A_1 f^B_2 (1-f^C_3)(1-f^D_4)]+\sum\limits_{B,C,D,E,F}\Bigg\{\int\limits_{2,3,4,5,6}\frac{\mathrm{d}p_5}{2\pi}\frac{\mathrm{d}p_6}{2\pi}\, \mathrm{d}\mathnormal{\Gamma}_{5 6\rightarrow 1 2 3 4}^{E,F;A,B,C,D} L\, S_{5 6\rightarrow (1) 2 3 4}^{E,F;A,B,C,D}\, \notag\\
&\times [f^E_5 f^F_6 (1-f^A_1)(1-f^B_2)(1-f^C_3)(1-f^D_4)-f^A_1 f^B_2 f^C_3 f^D_4 (1-f^E_5)(1-f^F_6)]\notag\\
&+\int\limits_{2,3,4,5,6}\frac{\mathrm{d}p_1}{2\pi}\frac{\mathrm{d}p_2}{2\pi}\, \mathrm{d}\mathnormal{\Gamma}_{1 2\rightarrow 3 4 5 6}^{A,B;C,D,E,F} L\, S_{(1) 2\rightarrow 3 4 5 6}^{A,B;C,D,E,F}\, [f^C_3 f^D_4 f^E_5 f^F_6 (1-f^A_1)(1-f^B_2)\notag\\
&-f^A_1 f^B_2 (1-f^C_3)(1-f^D_4)(1-f^E_5)(1-f^F_6)]+\int\limits_{2,3,4,5,6}\frac{\mathrm{d}p_1}{2\pi}\frac{\mathrm{d}p_2}{2\pi}\frac{\mathrm{d}p_3}{2\pi}\, \mathrm{d}\mathnormal{\Gamma}_{1 2 3\rightarrow 4 5 6}^{A,B,C;D,E,F} L^2\, S_{(1) 2 3\rightarrow 4 5 6}^{A,B,C;D,E,F}\notag\\
&\times [f^D_4 f^E_5 f^F_6 (1-f^A_1)(1-f^B_2)(1-f^C_3)-f^A_1 f^B_2 f^C_3 (1-f^D_4)(1-f^E_5)(1-f^F_6)]\,\Bigg\}\ ,\label{Eq:collision1}
\end{align}

\noindent where the sum over indices runs over all the possible configurations of fermion, anti-fermion, and flavor states. The symmetry factors (to be specified later) $S_{(1) 2\leftrightarrow 3 4}$, $S_{(1) 2\leftrightarrow 3 4 5 6}$, $S_{5 6\leftrightarrow (1) 2 3 4}$, and $S_{(1) 2 3\leftrightarrow 4 5 6}$ avoid the double counting from relabeling of momenta for identical particles (except for the particle denoted as `1') in the integral and considering equivalent processes after summing over all the fermion types. The transition rate for an arbitrary process $\alpha\rightarrow\beta$ is given in terms of the scattering amplitude $\mathcal{M}$ by \cite{Wei00,Pes95}
\begin{align}
\mathrm{d}\mathnormal{\Gamma}(\alpha\rightarrow\beta)=L^{1-N_\alpha}\left[\prod\limits_\alpha(2E_\alpha)^{-1}\right] \left[\prod\limits_\beta\frac{\mathrm{d}p_\beta}{(2\pi) 2 E_\beta}\right]\, |\mathcal{M}(\alpha\rightarrow\beta)|^2 (2\pi)^2\delta^{(2)}(\sum_\alpha\underline{p}{}_\alpha-\sum_\beta\underline{p}{}_\beta)\ ,
\end{align}
where $L$ is the size of the system (although we consider the limit $L\rightarrow\infty$), and $N_\alpha$ the number of particles in the initial state.

\vspace{0.3cm}In order to obtain an expression for the bulk viscosity we need to solve (\ref{Eq:Boltzmann}) for small departures from equilibrium. To do it, we first write
\begin{align}\label{Eq:fdecomp}
f^A(t,x,\underline{k})=f^A_\mathrm{eq}(t,x,\underline{k})+\delta f^A(t,x,\underline{k})\ ,
\end{align}
where $\delta f^A$ is small, and the fermion or anti-fermion distribution function at equilibrium for zero chemical potential is
\begin{align}
f^a_\mathrm{eq}(t,x,\underline{k})=\frac{1}{\mathrm{e}^{\beta \underline{k}\cdot u}+1}\ ,
\end{align}
with $\beta^{-1}\equiv T(t,x)$ the local temperature, and $u^\mu(t,x)$ the velocity of the corresponding fluid cell. Expanding the left-hand side of (\ref{Eq:Boltzmann}) in the local rest frame ($u^1|_\mathrm{l.r.f.}=0$) to linear order in spatial derivatives, we obtain\footnote{Here we make use of the thermodynamic relations $\mathrm{d}T/T=\mathrm{d}P/(\epsilon+P)$, $c_\mathrm{s}^2=\partial P/\partial\epsilon$.  Also, from the conservation of the energy-momentum tensor, $\partial_\mu T^{\mu\nu}=0$, applied to leading order to the perfect fluid, $T^{\mu\nu}_\mathrm{p.f.}=-P g^{\mu\nu}+(\epsilon+P)u^\mu u^\nu$, we derive the relations in the local rest frame ($\partial_\mu u^0|_\mathrm{l.r.f.}=0$): $\partial\epsilon/\partial t=-(\epsilon+P)\partial u^1/\partial x$, $\partial u^1/\partial t=-(\epsilon+P)^{-1}\partial P/\partial x$.}
\begin{align}\label{Eq:source}
\left(\frac{\partial}{\partial t}+\frac{k}{E_k}\frac{\partial}{\partial x}\right)f^A\Big|_\mathrm{l.r.f.}\simeq\ -n_\mathrm{F}(E_k)[1-n_\mathrm{F}(E_k)]\beta\left[\left(E_k-T\frac{M}{E_k}\frac{\mathrm{d}M}{\mathrm{d}T}\right) c_\mathrm{s}^2-\frac{k^2}{E_k}\right]\frac{\partial u^1}{\partial x}\ ,
\end{align}
where $c_\mathrm{s}$ is the speed of sound in the fluid.

Consequently, the deviation from equilibrium can be written in the form
\begin{align}\label{Eq:deviation}
\left.\delta f^A_k\right|_\mathrm{l.r.f.}=-\beta n_\mathrm{F}(E_k)[1-n_\mathrm{F}(E_k)]\mathcal{B}^A_k\,\frac{\partial u^1}{\partial x} ,
\end{align}
with $\mathcal{B}^A_k=\mathcal{B}^A(|k|)$ some dimensionless function to be determined by solving the integral equation obtained after the previous linearization
of both sides of (\ref{Eq:Boltzmann}):
\begin{align}
&p_1^2-c_\mathrm{s}^2 \left(E_1^2-TM\frac{\mathrm{d}M}{\mathrm{d}T}\right)=\frac{1}{2(1-n_\mathrm{F,1})}\Bigg\{\sum\limits_{B,C,D}\,\int\limits_{-\infty}^\infty
\left[\prod\limits_{i=2}^4\frac{\mathrm{d}p_i}{(2\pi)2E_i}\right]\, |\mathcal{M}_{A,B}^{C,D}(p_1,p_2;p_3,p_4)|^2\notag\\ &\times S_{(1) 2\rightarrow 3 4}^{A B; C D}(2\pi)^2\delta^{(2)}(\underline{p}{}_1+\underline{p}{}_2-\underline{p}{}_3-\underline{p}{}_4)n_\mathrm{F,2}(1-n_\mathrm{F,3}) (1-n_\mathrm{F,4})(\mathcal{B}^A_1+\mathcal{B}^B_2-\mathcal{B}^C_3-\mathcal{B}^D_4)\notag\\
&+\sum\limits_{B,C,D,E,F}\,\int\limits_{-\infty}^\infty\left[\prod\limits_{i=2}^6\frac{\mathrm{d}p_i}{(2\pi)2E_i}\right]
\Bigg\{\,|\mathcal{M}_{E,F}^{A,B,C,D}(p_5,p_6;p_1,p_2,p_3,p_4)|^2 S_{5 6\rightarrow (1) 2 3 4}^{E F; A B C D}\notag\\
&\times(2\pi)^2\delta^{(2)}(\underline{p}{}_5+\underline{p}{}_6-\underline{p}{}_1-\underline{p}{}_2-\underline{p}{}_3-\underline{p}{}_4)n_\mathrm{F,2}
n_\mathrm{F,3}n_\mathrm{F,4}(1-n_\mathrm{F,5})(1-n_\mathrm{F,6})\notag\\
&\times(\mathcal{B}^A_1+\mathcal{B}^B_2+\mathcal{B}^C_3+\mathcal{B}^D_4-\mathcal{B}^E_5-\mathcal{B}^F_6)+|\mathcal{M}_{A,B}^{C,D,E,F} (p_1,p_2;p_3,p_4,p_5,p_6)|^2 S_{(1) 2\rightarrow 3 4 5 6}^{A B; C D E F} (2\pi)^2\notag\\
&\times\delta^{(2)}(\underline{p}{}_1+\underline{p}{}_2-\underline{p}{}_3-\underline{p}{}_4-\underline{p}{}_5-\underline{p}{}_6)n_\mathrm{F,2} (1-n_\mathrm{F,3})(1-n_\mathrm{F,4})(1-n_\mathrm{F,5})(1-n_\mathrm{F,6})\notag\\
&\times(\mathcal{B}^A_1+\mathcal{B}^B_2-\mathcal{B}^C_3-\mathcal{B}^D_4-\mathcal{B}^E_5-\mathcal{B}^F_6)+|\mathcal{M}_{A,B,C}^{D,E,F} (p_1,p_2,p_3;p_4,p_5,p_6)|^2 S_{(1) 2 3\rightarrow 4 5 6}^{A B C; D E F}\notag\\
&\times(2\pi)^2\delta^{(2)}(\underline{p}{}_1+\underline{p}{}_2+\underline{p}{}_3-\underline{p}{}_4-\underline{p}{}_5-\underline{p}{}_6)n_\mathrm{F,2} n_\mathrm{F,3}(1-n_\mathrm{F,4})(1-n_\mathrm{F,5})(1-n_\mathrm{F,6})\notag\\
&\times(\mathcal{B}^A_1+\mathcal{B}^B_2+\mathcal{B}^C_3-\mathcal{B}^D_4-\mathcal{B}^E_5-\mathcal{B}^F_6)\Bigg\}\Bigg\}\ .\label{Eq:integral1}
\end{align}

We can interpret the right-hand side of (\ref{Eq:integral1}) as the action of a linear operator $\hat{\mathscr{C}}$ over a function in the space of solutions of the transport equation, and we split this operator into two terms, $\hat{\mathscr{C}}\equiv\hat{\mathscr{C}}_\mathrm{el}+\hat{\mathscr{C}}_\mathrm{in}$, corresponding to elastic and inelastic processes respectively. At this point, an important simplification is in order. Since the source term in (\ref{Eq:source}) is invariant under charge conjugation, and the theory is symmetric under $O(N)$-flavor rotations (this symmetry cannot be broken in $1+1$ dimensions \cite{Col73}), then the departures from equilibrium are the same for all the particle types, i.e., $\mathcal{B}^A_k\equiv \mathcal{B}(|k|)$. Furthermore, the $\delta$-function in the elastic $2\rightarrow 2$ term of the collision integral implies in $1+1$ dimensions that the final set of momenta are the same as the initial, i.e., $2\rightarrow 2$ elastic collisions in $1+1$ dimensions cannot relax back to equilibrium a perturbation in the distribution of momenta. Thus, $\hat{\mathscr{C}}_\mathrm{el}(2\rightarrow 2)=\hat{0}$ and therefore, to leading order in the large-$N$ expansion, $\hat{\mathscr{C}}=\hat{\mathscr{C}}_\mathrm{el}(3\rightarrow 3)+\hat{\mathscr{C}}_\mathrm{in}(2\leftrightarrow 4)$.

\vspace{0.3cm}Once we know $\mathcal{B}(|k|)$, from (\ref{Eq:deviation}), (\ref{Eq:const11}), and (\ref{Eq:emtensor}) it is straightforward to obtain the bulk viscosity:\footnote{The Landau-Lifshitz condition
\begin{align}
0=\int\limits_{-\infty}^\infty\frac{\mathrm{d}k}{2\pi E_k}\, (E_k^2-TM\mathrm{d}M/\mathrm{d}T) n_\mathrm{F}(E_k)[1-n_\mathrm{F}(E_k)]\mathcal{B}^A(\underline{k})\ ,
\end{align}
imposed to make the decomposition (\ref{Eq:fdecomp}) unique, is also used here \cite{Jeo96,Cha10}.}
\begin{align}
\zeta =\beta\sum\limits_{A}\int\limits_{-\infty}^\infty\frac{\mathrm{d}k}{2\pi E_k}\, n_\mathrm{F}(E_k)[1-n_\mathrm{F}(E_k)] \left[k^2-c_\mathrm{s}^2\left(E_k^2-TM\frac{\mathrm{d}M}{\mathrm{d}T}\right)\right] \mathcal{B}^A(\underline{k})\ .\label{Eq:bulk1}
\end{align}

The linearized version of the Boltzmann equation (\ref{Eq:integral1}) can be written as
\begin{align}\label{Eq:Boltzmannkets}
|\mathcal{S}\rangle=\hat{\mathscr{C}}|\mathcal{B}\rangle\ ,
\end{align}
where $\mathcal{S}(\underline{p})\equiv p^2-c_\mathrm{s}^2 (E_p^2-T M\, \mathrm{d}M/\mathrm{d}T)$ denotes the source term. Defining the scalar product of two square-integrable functions as
\begin{align}
\langle\chi|\psi\rangle\equiv\beta\sum\limits_A\int\limits_{-\infty}^\infty\frac{\mathrm{d}k}{(2\pi)E_k}\, n_\mathrm{F}(E_k)[1-n_\mathrm{F}(E_k)]\,\chi^A(\underline{k})\psi^A(\underline{k})\ ,
\end{align}
then the bulk viscosity is given by
\begin{align}
\zeta=\langle\mathcal{S}|\mathcal{B}\rangle=\langle\mathcal{S}|\hat{\mathscr{C}}^{-1}|\mathcal{S}\rangle\ .
\end{align}
As shown in \cite{Arn00,Arn03,Arn06a}, in order to calculate numerically this expectation value for the inverse of the collision operator, it is optimal to do it variationally. If we define the functional
\begin{align}
Q[\chi]\equiv\langle\chi|\mathcal{S}\rangle-\frac{1}{2}\langle\chi|\hat{\mathscr{C}}|\chi\rangle\ ,
\end{align}
then the solution of (\ref{Eq:Boltzmannkets}) corresponds to a maximum in this functional, $\left.\delta Q/\delta \chi\right|_{\chi=\mathcal{B}}=0$. Hence, the bulk viscosity is proportional to this maximum:
\begin{align}
\zeta=2\,Q_\mathrm{max}\ .
\end{align}
We now expand the solution for the Boltzmann equation in terms of a given set of $n$ linearly independent functions:
\begin{align}
\mathcal{B}(k)=\sum\limits_{i=1}^n b_i\phi_i(k)\ .
\end{align}
Then
\begin{align}\label{Eq:Qcoeff}
Q[\{b_i\}]=\sum\limits_{i=1}^n b_i S_i-\frac{1}{2}\sum\limits_{i,j=1}^n b_i C_{ij} b_j\ ,
\end{align}
where
\begin{align}
S_i\equiv\langle\phi_i|\mathcal{S}\rangle\ ,\quad C_{ij}\equiv\langle\phi_i|\hat{\mathscr{C}}|\phi_j\rangle\ .
\end{align}
Maximizing (\ref{Eq:Qcoeff}) with respect to the set of coefficients $\{b_i\}$, implies $\tilde{b}=\tilde{C}^{-1}\tilde{S}$ (a tilde denotes matrices), and therefore
\begin{equation}
\zeta=\tilde{S}^\mathrm{t}\tilde{b}=\tilde{S}^\mathrm{t}\tilde{C}^{-1}\tilde{S}\ .
\end{equation}
It is important to notice, from (\ref{Eq:integral1}), that the collision operator has one zero-mode $\chi_\mathrm{e}(\underline{p})\equiv E_p$ corresponding to energy conservation, i.e., $\hat{\mathscr{C}}|\chi_\mathrm{e}\rangle=0$.\footnote{\label{Ft:lowTinel}In addition to $\chi_\mathrm{e}$, the elastic collision operator also has the zero-mode $\chi_\mathrm{n}(|k|)\equiv 1$ corresponding to the conservation of the total number of particles in these type of processes. However, this is not a zero-mode of the inelastic part of the collision integral and therefore we do not have to worry about it when calculating $\hat{\mathscr{C}}^{-1}$. The presence of this other zero-mode, though, implies that the bulk viscosity is dominated by inelastic processes at very low temperatures due to Fermi-Dirac factors:
\begin{equation}\label{Eq:zerondominance}
\zeta=\langle\mathcal{S}|\hat{\mathscr{C}}^{-1}|\mathcal{S}\rangle\sim\frac{|\langle\chi_\mathrm{n}|\mathcal{S}\rangle|^2}{\langle\chi_\mathrm{n}|\hat{\mathscr{C}}_\mathrm{in}(2\leftrightarrow 4)|\chi_\mathrm{n}\rangle}\ ,\quad\text{for}\ T\ll M_0\ .
\end{equation}
On the other hand, at temperatures close to $T_\mathrm{c}$, the Fermi-Dirac factors are $\mathcal{O}(1)$ and both $3\rightarrow 3$ and $2\leftrightarrow 4$ processes are equally important.
} Therefore, in order to be able to invert the collision matrix, it is necessary to calculate it in the vector space orthogonal to this zero-mode. This does not affect the result for the bulk viscosity because the source term is orthogonal to this zero-mode, $\langle \mathcal{S}|\chi_\mathrm{e}\rangle=0$.

Besides the simplifications already commented in the previous paragraphs, in order to obtain an explicit expression for the matrix element of the collision operator in the large-$N$ limit, it is obvious (cf. Appendix \ref{App:amplitudes}) that the dominant scattering processes are those for which three different flavors participate. Moreover, the symmetry factors have to be specified, see Table \ref{Tab:symmetryf}.

\begin{table}[h!]

{\small
\begin{tabular}{|c|c|}\hline
$3\rightarrow 3$ processes&$S_{(1) 2 3\rightarrow 4 5 6}$\\\hline\hline
$\mathrm{f}^a\mathrm{f}^b\mathrm{f}^c\rightarrow\mathrm{f}^a\mathrm{f}^b\mathrm{f}^c$&1/12\\\hline
$\mathrm{f}^a\bar{\mathrm{f}}^a\mathrm{f}^b$, $\bar{\mathrm{f}}^a\mathrm{f}^a\mathrm{f}^b\rightarrow\bar{\mathrm{f}}^c\mathrm{f}^c\mathrm{f}^b$&1/12\\\hline
$\mathrm{f}^a\bar{\mathrm{f}}^a\mathrm{f}^c$, $\bar{\mathrm{f}}^a\mathrm{f}^a\mathrm{f}^c\rightarrow\bar{\mathrm{f}}^b\mathrm{f}^b\mathrm{f}^c$&1/12\\\hline
$\mathrm{f}^a\bar{\mathrm{f}}^b\mathrm{f}^b\rightarrow\bar{\mathrm{f}}^c\mathrm{f}^c\mathrm{f}^a$&1/12\\\hline
$\mathrm{f}^a\bar{\mathrm{f}}^c\mathrm{f}^c\rightarrow\bar{\mathrm{f}}^b\mathrm{f}^b\mathrm{f}^a$&1/12\\\hline
$\mathrm{f}^a\bar{\mathrm{f}}^b\mathrm{f}^c\rightarrow\mathrm{f}^a\bar{\mathrm{f}}^b\mathrm{f}^c$&1/12\\\hline
$\mathrm{f}^a\bar{\mathrm{f}}^c\mathrm{f}^b\rightarrow\mathrm{f}^a\bar{\mathrm{f}}^c\mathrm{f}^b$&1/12\\\hline
$\bar{\mathrm{f}}^a\mathrm{f}^b\mathrm{f}^c\rightarrow\bar{\mathrm{f}}^a\mathrm{f}^b\mathrm{f}^c$&1/12\\\hline
\end{tabular}
\hspace{0.5cm}\begin{tabular}{|c|c|c|}\hline
$2\leftrightarrow 4$ processes&$S_{(1) 2\rightarrow 3 4 5 6}$&$S_{5 6\rightarrow (1) 2 3 4}$\\\hline\hline
$\mathrm{f}^a\mathrm{f}^b\leftrightarrow\mathrm{f}^a\mathrm{f}^b\bar{\mathrm{f}}^c\mathrm{f}^c$&1/24&1/12\\\hline
$\mathrm{f}^a\mathrm{f}^c\leftrightarrow\mathrm{f}^a\mathrm{f}^c\bar{\mathrm{f}}^b\mathrm{f}^b$&1/24&1/12\\\hline
$\mathrm{f}^b\mathrm{f}^c\leftrightarrow\mathrm{f}^a\bar{\mathrm{f}}^a\mathrm{f}^b\mathrm{f}^c$, $\bar{\mathrm{f}}^a\mathrm{f}^a\mathrm{f}^b\mathrm{f}^c$&--&1/12\\\hline
$\bar{\mathrm{f}}^b\mathrm{f}^b\leftrightarrow\mathrm{f}^a\bar{\mathrm{f}}^a\bar{\mathrm{f}}^c\mathrm{f}^c$, $\bar{\mathrm{f}}^a\mathrm{f}^a\bar{\mathrm{f}}^c\mathrm{f}^c$&--&1/24, 1/12\\\hline
$\bar{\mathrm{f}}^c\mathrm{f}^c\leftrightarrow\mathrm{f}^a\bar{\mathrm{f}}^a\bar{\mathrm{f}}^b\mathrm{f}^b$, $\bar{\mathrm{f}}^a\mathrm{f}^a\bar{\mathrm{f}}^b\mathrm{f}^b$&--&1/24, 1/12\\\hline
$\bar{\mathrm{f}}^b\mathrm{f}^c\leftrightarrow\mathrm{f}^a\bar{\mathrm{f}}^a\bar{\mathrm{f}}^b\mathrm{f}^c$, $\bar{\mathrm{f}}^a\mathrm{f}^a\bar{\mathrm{f}}^b\mathrm{f}^c$&--&1/24, 1/12\\\hline
$\bar{\mathrm{f}}^c\mathrm{f}^b\leftrightarrow\mathrm{f}^a\bar{\mathrm{f}}^a\bar{\mathrm{f}}^c\mathrm{f}^b$, $\bar{\mathrm{f}}^a\mathrm{f}^a\bar{\mathrm{f}}^c\mathrm{f}^b$&--&1/24, 1/12\\\hline
$\mathrm{f}^a\bar{\mathrm{f}}^b\leftrightarrow\mathrm{f}^a\bar{\mathrm{f}}^b\bar{\mathrm{f}}^c\mathrm{f}^c$&1/48&1/24\\\hline
$\mathrm{f}^a\bar{\mathrm{f}}^c\leftrightarrow\mathrm{f}^a\bar{\mathrm{f}}^b\bar{\mathrm{f}}^c\mathrm{f}^b$&1/48&1/24\\\hline
$\bar{\mathrm{f}}^a\mathrm{f}^b\leftrightarrow\bar{\mathrm{f}}^a\bar{\mathrm{f}}^c\mathrm{f}^c\mathrm{f}^b$&1/48&1/12\\\hline
$\bar{\mathrm{f}}^a\mathrm{f}^c\leftrightarrow\bar{\mathrm{f}}^a\bar{\mathrm{f}}^b\mathrm{f}^b\mathrm{f}^c$&1/48&1/12\\\hline
$\mathrm{f}^a\bar{\mathrm{f}}^a$, $\bar{\mathrm{f}}^a\mathrm{f}^a\leftrightarrow\bar{\mathrm{f}}^b\mathrm{f}^b\bar{\mathrm{f}}^c\mathrm{f}^c$&1/48&--\\\hline
\end{tabular}}
\caption{Symmetry factors corresponding to relabeling the momenta of identical particles under the collision integral after summing over all particle types. Here $a\neq b\neq c\neq a$ denote only flavor (processes with three different flavors dominate). The momenta for each particular process are initially labeled according to the order indicated in the title of columns 2 and 3. The particle labeled with `(1)' is ``distinguishable''. Some cells are empty because the corresponding process has already been taken into account by another symmetry factor. The processes obtained by permutation of the three flavors, although omitted in the table, have the same symmetry factors and also have to be taken into account.}\label{Tab:symmetryf}
\end{table}

Finally,
\begin{align}\label{Eq:Cmatrix}
C_{ij}\simeq &\, N^3\beta\int\limits_{-\infty}^\infty
\left[\prod\limits_{i=1}^6\frac{\mathrm{d}p_i}{(2\pi)2E_i}\right]\,\Big\{(2\pi)^2\delta^{(2)}
(\underline{p}{}_1+\underline{p}{}_2-\underline{p}{}_3-\underline{p}{}_4-\underline{p}{}_5-\underline{p}{}_6) n_\mathrm{F,1}n_\mathrm{F,2}\notag\\
&\times(1-n_\mathrm{F,3})(1-n_\mathrm{F,4})(1-n_\mathrm{F,5})(1-n_\mathrm{F,6})\left[|\mathcal{M}_{12\rightarrow \bar{3}456}|^2+\frac{3}{2}\,|\mathcal{M}_{\bar{1}2\rightarrow \bar{3}\bar{4}56}|^2\right]\notag\\
&\times [\phi_i(p_1)+\phi_i(p_2)-\phi_i(p_3)-\phi_i(p_4)-\phi_i(p_5)-\phi_i(p_6)]\notag\\
&\times [\phi_j(p_1)+\phi_j(p_2)-\phi_j(p_3)-\phi_j(p_4)-\phi_j(p_5)-\phi_j(p_6)]\notag\\
&+(2\pi)^2\delta^{(2)}
(\underline{p}{}_1+\underline{p}{}_2+\underline{p}{}_3-\underline{p}{}_4-\underline{p}{}_5-\underline{p}{}_6) n_\mathrm{F,1}n_\mathrm{F,2}n_\mathrm{F,3}(1-n_\mathrm{F,4})(1-n_\mathrm{F,5})(1-n_\mathrm{F,6})\notag\\
&\times \left[\frac{1}{6}\,|\mathcal{M}_{123\rightarrow 456}|^2+\frac{3}{2}\,|\mathcal{M}_{\bar{1}23\rightarrow \bar{4}56}|^2\right] [\phi_i(p_1)+\phi_i(p_2)+\phi_i(p_3)-\phi_i(p_4)-\phi_i(p_5)-\phi_i(p_6)]\notag\\
&\times [\phi_j(p_1)+\phi_j(p_2)+\phi_j(p_3)-\phi_j(p_4)-\phi_j(p_5)-\phi_j(p_6)]\Big\}\ ,
\end{align}
where $\mathcal{M}_{12\rightarrow \bar{3}456}$ and $\mathcal{M}_{\bar{1}2\rightarrow \bar{3}\bar{4}56}$ are the inelastic amplitudes of fermion-fermion and anti-fermion-fermion scattering respectively. The amplitude squared $|\mathcal{M}_{\bar{1}\bar{2}\rightarrow \bar{3}\bar{4}\bar{5}6}|^2$ of anti-fermion-anti-fermion scattering, as well as its corresponding symmetry factor, are equal to the fermion-fermion ones by charge conjugation, and have already been included in (\ref{Eq:Cmatrix}).

It is evident from (\ref{Eq:Cmatrix}) that the collision matrix is symmetric and positive semidefinite (positive definite in the space orthogonal to its only zero-mode). Note also that $S_i=\mathcal{O}(N)$, and since $C_{ij}=\mathcal{O}(1/N)$ (cf. Appendix \ref{App:amplitudes}), therefore $\zeta=\mathcal{O}(N^3)$.
\subsection{Numerical results}
A particularly convenient set of functions, which becomes a basis when $n\rightarrow\infty$, is \cite{Arn06a}
\begin{align}\label{Eq:basis}
\phi_i(k)=\frac{(|k|/\langle|k|\rangle)^{i-1}}{(1+|k|/\langle|k|\rangle)^{n-3}}\ ,\quad\ i=1,\dots,n\ ,
\end{align}
with the thermal average $\langle|k|\rangle\sim\sqrt{M_0 T}$ for $T\rightarrow 0$, $\langle|k|\rangle\sim T$ for $T\rightarrow \infty$, and interpolating between these two behaviors for intermediate temperatures. This set of functions automatically incorporates the required asymptotic behavior for the solution of the Boltzmann equation in the bulk channel: $B(|k|)\sim 1$ for $|k|\rightarrow 0$, and $B(|k|)\sim k^2$ for $|k|\rightarrow \infty$.

\begin{figure}[h!]
\centerline{\includegraphics[width=10cm]{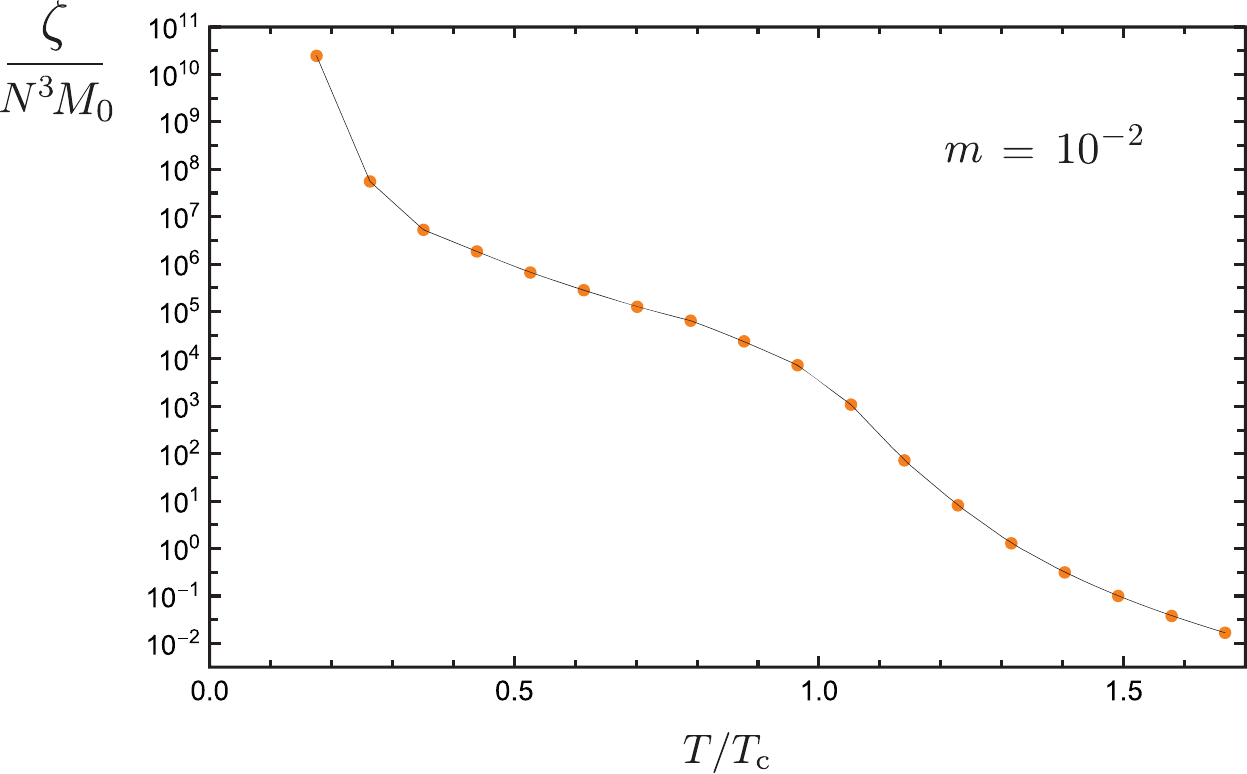}}
\caption{Bulk viscosity of the massive Gross-Neveu model for $m=10^{-2}M_0$, calculated with $n=3$ basis functions. The continuous line simply joins the data points.}\label{Fig:bulkv}
\end{figure}

In Fig. \ref{Fig:bulkv}, I plot the numerical result of a variational computation of the bulk viscosity in the massive Gross-Neveu model with $m=10^{-2} M_0$ using $n=3$ basis functions. It is not difficult to realize that $\zeta$ increases exponentially at low temperatures, like $\sim\exp(2M_0/T$), due to the Fermi-Dirac factors present in $\tilde{C}$ as well as in $\tilde{S}$ (cf. footnote \ref{Ft:lowTinel}). This behavior is analogous to the case of $\zeta$ for $\lambda\phi^4$ in $3+1$ dimensions \cite{Jeo95}.

\begin{figure}[h!]
\centerline{\includegraphics[width=11cm]{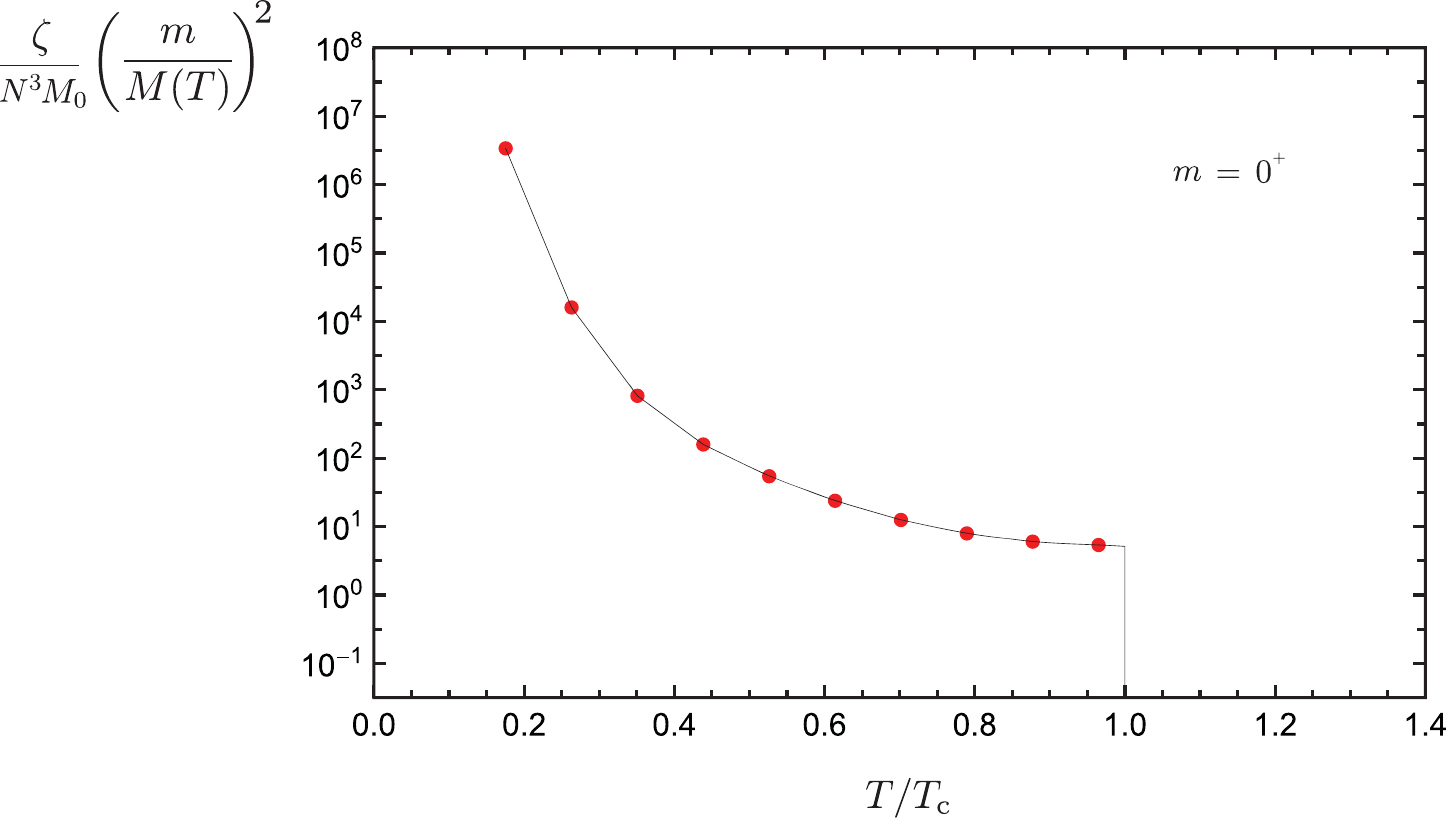}}
\caption{Bulk viscosity of the massive Gross-Neveu model for $m=0^+$, calculated with $n=3$ basis functions. The continuous line simply joins the data points.}\label{Fig:bulkv0}
\end{figure}

The numerical error corresponding to considering only $n=3$ basis functions (including the error from the numerical evaluation of integrals) is estimated to be of order $0.5\%$ for temperatures around $T_\mathrm{c}$, increasing as we go down in temperatures, being $\sim 60\%$ for $T=0.1 M_0$, which indicates that the basis (\ref{Eq:basis}) is not the best choice at those temperatures. By considering $n=9$, the precision can be improved to $\sim 20\%$ at $T=0.1 M_0$. However, due to the exponential growth of $\zeta$ at low $T$ and since the result shown corresponds to a lower bound, the qualitative behavior with temperature is not expected to change significantly.
From the numerical result we clearly see that there is no maximum in the bulk viscosity near $T_\mathrm{c}$, it is a monotonously decreasing function of the temperature. By reducing further the value of $m$, we would eventually reconstruct (continuously) a discontinuity for $\zeta$ at $T_\mathrm{c}$. For infinitesimally small $m$, above $T_c$ the bulk viscosity would be arbitrarily small. Going down in temperatures, it would increase very sharply right at $T_\mathrm{c}$ (with an arbitrarily large value\footnote{This is perfectly fine for our purpose of testing the possible correlation between the bulk viscosity and the trace anomaly; nonetheless, this model is not suitable to obtain for instance an estimate of the absolute value of the quotient $\zeta/s$ for QCD.}), and it would continue increasing exponentially at very low temperatures. This is shown in the plot of Fig. \ref{Fig:bulkv0}.

%

\section{Discussion}\label{Sec:discussion}

\subsection{Sum rule in the bulk channel}
\begin{figure}[h!]
\centerline{\includegraphics[width=3cm]{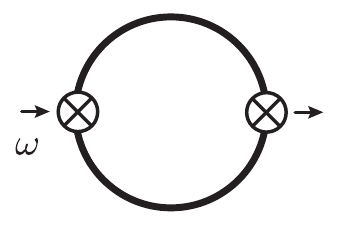}}
\caption{One of the diagrams contributing to $\delta\rho_\mathrm{bulk}/\omega$ at leading-order for $\omega\sim M_0$. The crosses denote insertions of the trace of the energy-momentum tensor. The fermion propagator is not dressed.}\label{Fig:spectraldensN}
\end{figure}

In the paper \cite{Rom09} (Section IV), the authors obtained a sum rule for the spectral density of the two-point function involving the trace of the energy-momentum tensor in pure Yang-Mills theory. The derivation is essentially based on the asymptotically-free character of the theory\footnote{However, there is a recent example of an asymptotically-free model for which the sum rule in the bulk channel has a different form from the one derived in \cite{Rom09}, due to the fact that conformal symmetry is not restored at high energies or temperatures in this model \cite{Spr10}. Note instead that in the massive Gross-Neveu model, although $m$ explicitly breaks scale symmetry, it is eventually restored at high energies because $m$ appears in the lagrangian multiplied by $g^2N$.}. In this subsection I am going to comment on an interesting aspect of the massive Gross-Neveu model concerning its sum rule in the bulk channel\footnote{A more rigorous and detailed analysis of the sum rule and the spectral density is underway and will be published elsewhere.}. Naively, since the massive Gross-Neveu model is asymptotically free, we can follow the analysis in \cite{Rom09} and convince ourselves that the version of the sum rule for this particular system is simply
\begin{align}\label{Eq:sumrule}
(\epsilon+P)(1-c_\mathrm{s}^2)-2(\epsilon-P)=\frac{2}{\pi}\int\limits_0^\infty\mathrm{d}\omega\, \frac{\delta\rho^\mathrm{bulk}(\omega)}{\omega}\ .
\end{align}
Interestingly, we notice that this sum rule does not depend explicitly on the mass parameter $m$. Since $\epsilon,P=\mathcal{O}(N)$ and $c_\mathrm{s}^2=\mathcal{O}(N^0)$, the LHS of (\ref{Eq:sumrule}) is $\mathcal{O}(N)$. On the other hand, the order of $\delta\rho_\mathrm{bulk}/\omega$ depends on the frequency. For frequencies of order $\omega\sim M_0$, a diagram contributing to $\delta\rho_\mathrm{bulk}/\omega$ at leading order is the one of Fig. \ref{Fig:spectraldensN}, which is $\mathcal{O}(N)$. Hence, the contribution to the integral in (\ref{Eq:sumrule}) from this region of frequencies is $\mathcal{O}(N)$, consistent with the LHS of the sum rule.

\begin{figure}[h!]
\centerline{\includegraphics[width=6cm]{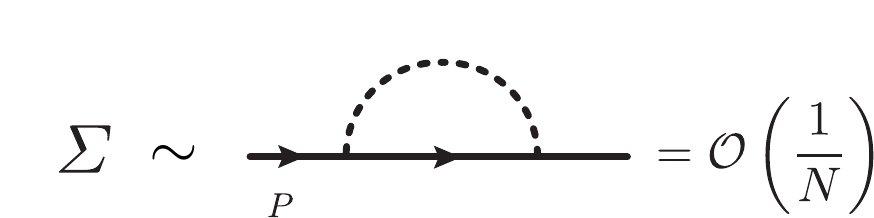}}
\caption{Leading-order contribution to the fermion self-energy.}\label{Fig:fermionwidth}
\end{figure}

\begin{figure}[h!]
\centerline{\includegraphics[width=6cm]{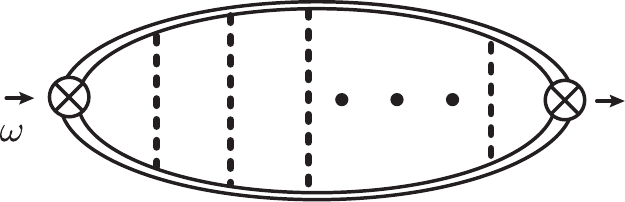}}
\caption{A ladder diagram with an arbitrary number of rungs contributing to $\delta\rho_\mathrm{bulk}/\omega$ at leading order for frequencies $\omega\sim 1/N$. Double lines denote dressing of the fermion propagator according to the diagram of Fig. \ref{Fig:fermionwidth}. The presence of pinching singularities is crucial in this regime of low frequencies.}\label{Fig:ladder}
\end{figure}

If we now consider lower frequencies, $\omega\sim 1/N$, then a resummation of diagrams is necessary to obtain the leading-order spectral density. This is essentially because of the presence of pinching singularities when the external frequency becomes of the order of the fermion width $\gamma_\mathrm{F}\sim\Im\mathnormal{\Sigma}=\mathnormal{O}(1/N)$ (cf. Fig. \ref{Fig:fermionwidth}) \cite{Jeo95,Val02,Aar02,Aar04,Aar05a}, or smaller. These singularities correspond to the product of retarded and advanced fermion propagators sharing approximately the same momentum:
\begin{align}
S_\mathrm{ret}(P)S_\mathrm{adv}(P)\sim\frac{1}{\gamma_\mathrm{F}}=\mathcal{O}(N)\ .
\end{align}
Consequently, the set of ladder diagrams depicted in Fig. \ref{Fig:ladder} all contribute at leading order, $\mathcal{O}(N^2)$, to the spectral density in this range of frequencies. Thus, their contribution to the integral is again consistent with the LHS of the sum rule.


For even lower frequencies, we know from the analysis of the Boltzmann equation in the previous section that $3\rightarrow 3$ and $2\leftrightarrow 4$ processes will eventually dominate the spectral density. Since the bulk viscosity is of order $\mathcal{O}(N^3)$, for frequencies close to zero $\delta\rho_\mathrm{bulk}/\omega=\mathcal{O}(N^3)$, and therefore, in order for it to be consistent with the LHS of the sum rule, this region of frequencies must be $\omega\lesssim 1/N^2$. Now, in the previous section we saw that for temperatures $T<T_\mathrm{c}$, if we reduce the value of $m$, the bulk viscosity can become arbitrarily large. However, the LHS of (\ref{Eq:sumrule}) remains finite as $m\rightarrow 0^+$, this implies that the region where $\delta\rho_\mathrm{bulk}/\omega$ is of order $\mathcal{O}(N^3)$ has a width $\sim 1/N^w$ with $w>2$ and therefore the bulk viscosity does not contribute to the sum rule in this regime of temperatures for any value of $m$.

\subsection{Other systems}

From the analysis of the previous sections for the massive Gross-Neveu model we can already extract some conclusions for other similar systems. Consider for instance the non-linear $\sigma$-model in $1+1$ dimensions \cite{Pol75}. This model also shares with massless QCD the features of asymptotic freedom, dynamical generation of a mass gap, and classical scale invariance broken by the trace anomaly. The lagrangian of the model is
\begin{align}
\mathcal{L}=\frac{1}{2}\,\partial_\mu\phi_a\partial^\mu\phi_a\ ,\quad a=1,\ldots,N\quad\text{with the condition}\quad \phi_a\phi_a=1/g^2\ .
\end{align}
Since the $O(N)$ symmetry cannot be broken in $1+1$ dimensions, there is no phase transition in this system at finite temperature. It again is convenient to introduce an auxiliary field $\alpha$ in order to analyze diagrammatically the large-$N$ limit:
\begin{align}
\mathcal{L}=\frac{1}{2}\partial_\mu\phi_a\partial^\mu\phi_a-\frac{1}{2}\mathrm{i}\alpha(\phi_a\phi_a-1/g^2)\ .
\end{align}
This model is also integrable \cite{Zam79}. To leading order in the $1/N$ expansion, the inelastic diagrams have the same topology as the ones in Fig. \ref{Fig:leadinginel}. Integrability is proven in an analogous way to the Gross-Neveu model using the factorization of scalar loops in $1+1$ dimensions. And it can be broken for instance by introducing a term $\sim \kappa(\phi_a\phi_a)^2$ in the lagrangian, so the correlation between the bulk viscosity and the trace anomaly can be studied by making $\kappa$ arbitrarily small.

\begin{figure}[h!]
\centerline{\includegraphics[width=16cm]{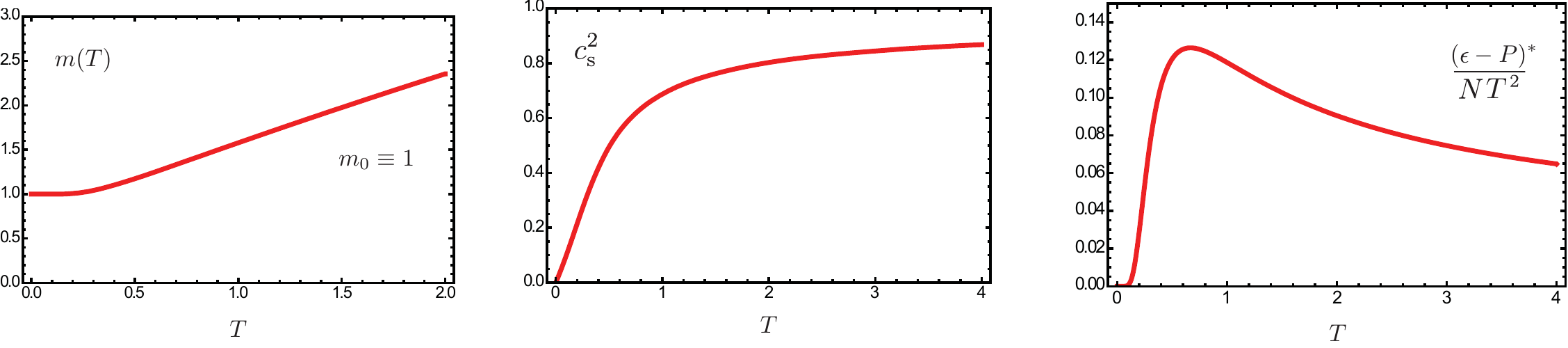}}
\caption{Thermal mass gap, speed of sound, and trace anomaly of the non-linear $\sigma$-model in $1+1$ dimensions to leading order in the $1/N$ expansion. $(\cdot)^*$ denotes the finite-temperature part.}\label{Fig:thermonlsm}
\end{figure}

\begin{figure}[h!]
\centerline{\includegraphics[width=11cm]{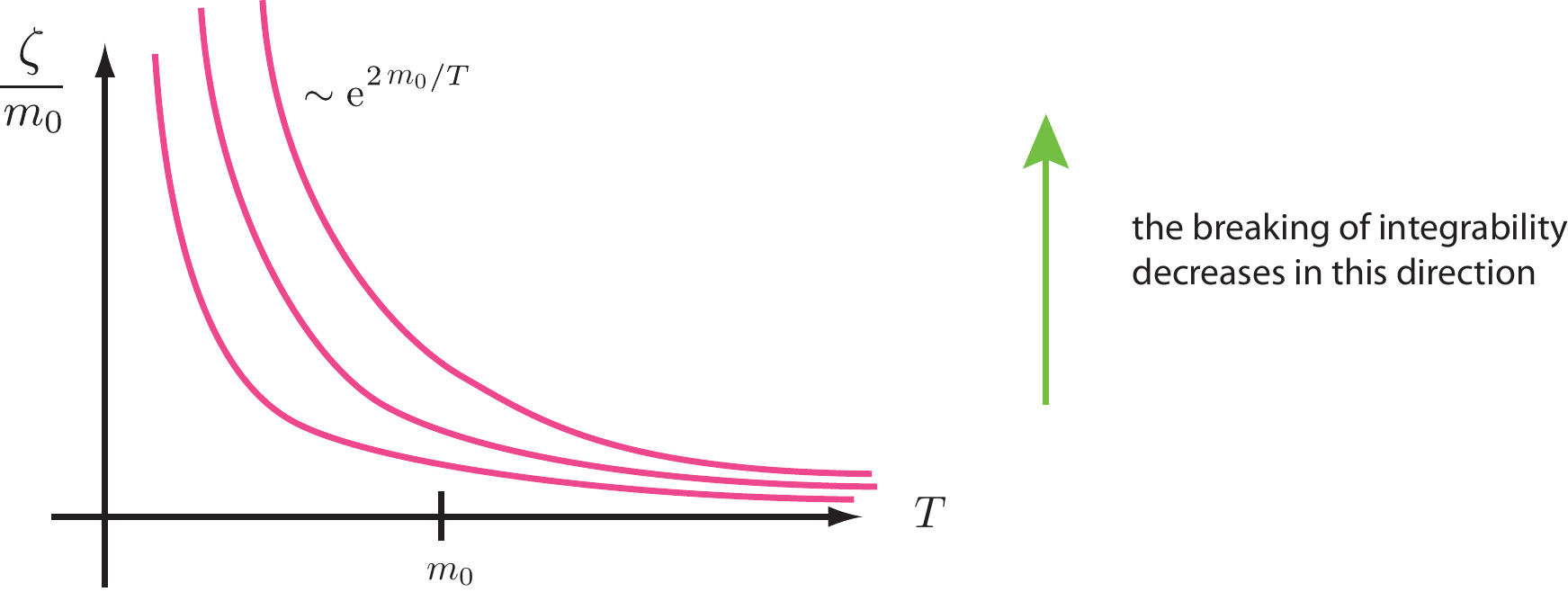}}
\caption{Qualitative behavior for the bulk viscosity of the non-linear $\sigma$-model in $1+1$ dimensions in the large-$N$ limit.}\label{Fig:bulkvnlsm}
\end{figure}

The thermodynamic properties of the system have been studied for instance in the works \cite{And04,And04a}. In Fig. \ref{Fig:thermonlsm} I plot the thermal mass gap, speed of sound, and trace anomaly. There are no discontinuities for these quantities in the limit $\kappa\rightarrow 0^+$. We then realize that the qualitative behavior of the bulk viscosity as we decrease $\kappa$ (restoring integrability) is the one depicted in Fig. \ref{Fig:bulkvnlsm}. For this model, integrability is always restored as $\kappa\rightarrow 0^+$ and therefore bulk viscosity diverges. One could also consider the sum rule, which presumably is identical to the case of the Gross-Neveu model, and similarly the bulk viscosity would not contribute to it.


Pure Yang-Mills theory in $3+1$ dimensions in the large-$N_\mathrm{c}$ limit is also very similar to the massive Gross-Neveu model regarding the bulk viscosity. For low energies and temperatures, interaction between glueballs is suppressed by powers of $N_\mathrm{c}$, being the 3-point scattering amplitude $\sim 1/N_\mathrm{c}$ \cite{Col85}. In this case the bulk viscosity is dominated by inelastic processes, due to the presence of a zero-mode in the collision integral corresponding to particle-number conservation \cite{Jeo95,Jeo96}. This implies an exponential growth of the bulk viscosity as the temperature decreases, $\sim\exp(m_\mathrm{g}/T)$, with $m_\mathrm{g}$ the mass of the lightest glueball. Regarding the sum rule derived in \cite{Rom09} for this theory, at temperatures below $T_\mathrm{c}$, the region of frequencies where inelastic processes dominate becomes very narrow to be consistent with the sum rule. This implies for instance that extracting transport coefficients on the lattice at temperatures below $T_\mathrm{c}$ becomes much more difficult as $N_\mathrm{c}$ increases \cite{Aar02}. In this regime of temperatures, a kinetic theory approach instead is more suitable in the large-$N_\mathrm{c}$ limit to obtain transport coefficients.

On the other hand, massless $N_\mathrm{f}=2$ QCD is qualitatively different from the previous models. At very low temperatures, the dynamics is dominated by Goldstone bosons and there is no exponential growth in the bulk viscosity as the temperature decreases. In addition, since this system undergoes a second-order phase transition in $3+1$ dimensions, the bulk viscosity would diverge at the critical temperature \cite{Pae06,Moo08,Kar08,Cha10}.

\section{Conclusions}\label{Sec:conclusions}

We have seen that the massive Gross-Neveu model is non-integrable in the large-$N$ limit, which allows the study of momentum transport in this system. We found that there is no direct correlation between the trace anomaly and the bulk viscosity in general, i.e., a peak in the former does not necessarily imply a peak in the latter.\footnote{Of course these two quantities are not independent of each other; conformal theories have zero bulk viscosity.} This was already obtained by S. Jeon in \cite{Jeo95} where he analyzed the bulk viscosity of massive $\lambda\phi^4$ theory in $3+1$ dimensions, but since it is not an asymptotically-free theory and the scale symmetry is explicitly broken in that case, it remained to analyze whether a QCD-like theory could be qualitatively different due to the anomaly (idea originally motivated by the paper \cite{Kha08}). The use of this simple model in $1+1$ dimensions also avoids interference from critical phenomena present in higher dimensions (for instance, the bulk viscosity would diverge near a second-order phase transition). In addition, it is a useful model to study sum rules and transport coefficients both at zero and finite fermion density in the large-$N$ limit. For instance, regarding sum rules, after a first superficial analysis, we saw that below $T_\mathrm{c}$ the bulk viscosity would not contribute to the sum rule. This implies in general that it is not necessarily possible to extract bulk viscosity from sum rules. Further work in these directions is in progress.

\begin{acknowledgments}
I thank Harmen Warringa for proposing to analyze the bulk viscosity of the non-linear sigma model in 1+1 dimensions, and for helpful discussions. I also want to thank useful discussions with Antonio Dobado, Kenji Fukushima, Angel Gomez Nicola, Elena Gubankova, Yoshimasa Hidaka, Xu-Guang Huang, and Toru Kojo.

This work has been sponsored by the Helmholtz International Center for FAIR. I acknowledge partial financial support from the Spanish research projects FIS2008-01323, FPA2008-00592, and UCM-BSCH GR58/08 910309.
\end{acknowledgments}

\appendix

\section{Factorization of fermion loops in $\maybebm{1+1}$ dimensions}\label{App:fermions}
In $1+1$ dimensions, the Lorentz group consists only of boots and therefore fermions have no spin. A two-dimensional representation of the Dirac algebra is for instance
\begin{align}
\gamma^0=\sigma^1=\Bigg(\begin{array}{ccc}0&&1\\1&&0\end{array}\Bigg)\ ,\quad \gamma^1=\mathrm{i}\sigma^2=\Bigg(\begin{array}{ccc}0&&1\\-1&&0\end{array}\Bigg)\quad \Rightarrow\quad \{\gamma^\mu,\gamma^\nu\}=2g^{\mu\nu}\ ,
\end{align}
with $\sigma^i$ the Pauli matrices, and $g=\diag(+1,-1)$.

The general solution of the Dirac equation $(\mathrm{i}\gamma^\mu\partial_\mu-m)\psi(x)=0$ can be written as a linear combination of plane waves
$u(p)\,\mathrm{e}^{-\mathrm{i}p\cdot x}\,,\, v(p)\,\mathrm{e}^{\mathrm{i}p\cdot x}$, corresponding to fermions and anti-fermions respectively, where $p^0>0$ and $p^2=m^2$. The spinors are normalized as
\begin{equation}\label{Eq:spinornorm}
\bar{u}(p)u(p)=2m\ ,\quad\bar{v}(p)v(p)=-2m\ ,
\end{equation}
and they also verify
\begin{equation}\label{Eq:spinorprod}
u(p)\bar{u}(p)=\psl +\, m\ ,\quad v(p)\bar{v}(p)=\psl -\,m\ .
\end{equation}

\begin{figure}[h!]
\begin{center}
\includegraphics[width=5cm]{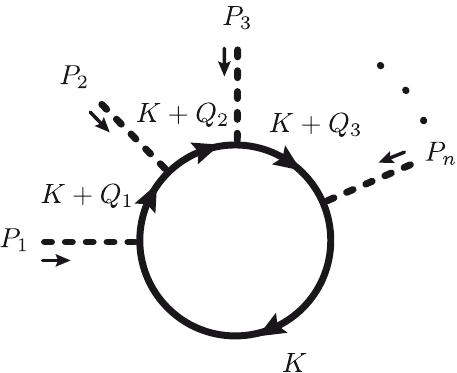}
\caption{A fermion loop with $n\geq 3$ external legs corresponding to the $\sigma$ field. The different momenta satisfy $Q_i=Q_{i-1}+P_i$, with $Q_0\equiv Q_n\equiv 0$, and $\sum_{i=1}^nP_i=0$.}\label{Fig:fermion-loop}
\end{center}
\end{figure}

In the rest of this Appendix, I will derive the finite-temperature version of the result previously obtained in \cite{Ber77} concerning the factorization of fermion loops in $1+1$ dimensions. Consider the momentum integral and Matsubara sum corresponding to the fermion loop in Fig. \ref{Fig:fermion-loop} at finite temperature with $n\geq 3$ (which is finite in $1+1$ dimensions):
\begin{align}
L_n&\equiv T\sum\limits_{\omega_m}\int\limits_{-\infty}^\infty\frac{\mathrm{d}k}{2\pi}\ \frac{M-(\Ksl+\Qsl_1)}{(K+Q_1)^2+M^2}\cdot\ldots\cdot\frac{M-(\Ksl+\Qsl_n)}{(K+Q_n)^2+M^2}\ ,\label{Eq:fermionL}
\end{align}
with $K=(-\omega_m,k)$, $\omega_m=(2m+1)\pi T$ ($m\in\mathbb{Z}$), $Q_i=(-\nu_i,q_i)$, $\nu_i=2r\pi T$ ($r\in\mathbb{Z}$).

We first perform the Matsubara sum using the result \cite{Val02}
\begin{align}
T\sum\limits_m F(\mathrm{i}\omega_m)=\sum\limits_{\mathrm{poles}\ z_i}n_\mathrm{F}(z_i)\Res(F;z_i)-\sum\limits_\mathrm{cuts}\int\limits_{-\infty}^\infty\frac{\mathrm{d}\xi}{2\pi\mathrm{i}}\ n_\mathrm{F}(\xi)\Disc (F;\mathrm{cut})\ .
\end{align}
The integrand in (\ref{Eq:fermionL}) has poles at $\mathrm{i}\omega_m=-\mathrm{i}\nu_i\pm E_{k+q_i}$, and no cuts. Thus,\footnote{In what follows I will be careless with Dirac indices, what makes manipulating expressions easier. However, we must somehow keep track of them so the final result has the right matrix structure in Dirac space.}
\begin{align}\label{Eq:Ln1}
L_n&=\int\limits_{-\infty}^\infty\frac{\mathrm{d}k}{2\pi}\,\sum\limits_{i=1}^n\sum\limits_{s=\pm 1} n_\mathrm{F}(s E_{k+q_i})\frac{M-\Ksl^s_{(i)}}{(-s)2 E_{k+q_i}}\prod\limits_{\tiny\begin{array}{c}j=1\\j\neq i\end{array}}^n\frac{M-(\Ksl^s_{(i)}+\Qsl_{ji})}{(K^s_{(i)}+Q_{ji})^2+M^2}\ ,
\end{align}
where $Q_{ji}\equiv Q_j-Q_i$, and $K^s_{(i)}\equiv (\mathrm{i}s E_{k+q_i},k+q_i)$ is on-shell, i.e., $(K^s_{(i)})^2=-M^2$. If we now make the change of variables
\begin{equation}\label{Eq:changev}
l_i\equiv\frac{k+q_i}{E_{k+q_i}-M}\ ,
\end{equation}
then the integrand in (\ref{Eq:Ln1}) becomes rational:
\begin{align}
L_n=&\ \frac{1}{2\pi}\sum\limits_{i=1}^n\left(\,\int\limits_{-\infty}^{-1}+\int\limits_1^\infty\,\right)\mathrm{d}l_i\, \sum\limits_{s=\pm 1}n_\mathrm{F}(s \tilde{E}_{(i)})\frac{M-\Ksl^s_{(i)}}{(-s)(l_i^2-1)}\notag\\
&\times\prod\limits_{\tiny\begin{array}{c}j=1\\j\neq i\end{array}}^n\frac{M-(\tilde{\Ksl}^s_{(i)}+\Qsl_{ji})}{\frac{1}{l_i^2-1}(\mathrm{i}2 M s q_{ji,2}+Q_{ji}^2)(l_i-l_{ji}^{+,s})(l_i-l_{ji}^{-,s})}\ ,
\end{align}
where $\tilde{E}_{(i)}$ and $\tilde{K}^s_{(i)}$ mean making the substitution $k\mapsto k(l_i,q_i)$ in $E_{k+q_i}$ and $K^s_{(i)}$ according to (\ref{Eq:changev}), explicitly
\begin{equation}
\tilde{E}_{(i)}=M\,\frac{l_i^2+1}{l_i^2-1}\ ,\quad \tilde{K}^s_{(i)}=\frac{M}{l_i^2-1}(\mathrm{i}s(l_i^2+1),2l_i)\ ,\label{Eq:tildeK}
\end{equation}
and $l_{ji}^{\pm,s}$ are the solutions of the equation
\begin{align}
(K^s_{(i)}+Q_{ji})^2+M^2=0\quad\Leftrightarrow\quad l_{i}=\frac{-2Mq_{ji,1}\pm|Q_{ji}^2|\sqrt{1+4M^2/Q_{ji}^2}}{\mathrm{i}2sMq_{ji,2}+Q_{ji}^2}\equiv l_{ji}^\pm\ .\label{Eq:zeros}
\end{align}
After partial fraction decomposition, we have\footnote{The terms corresponding to the poles at $l_i=\pm 1$ give a zero contribution to the integral (it can be easily seen from the fact that the integral must finite), so we ignore them.}
\begin{align}
L_n&=\sum\limits_{s=\pm 1}\sum\limits_{i=1}^n\left(\,\int\limits_{-\infty}^{-1}+\int\limits_1^\infty\,\right)\mathrm{d}l_i\ n_\mathrm{F}\left(s M\frac{l_i^2+1}{l_i^2-1}\right) \sum\limits_{\tiny\begin{array}{c}j=1\\j\neq i\end{array}}^n\left(\frac{\hat{A}_{ji}^{+,s}}{l_i-l_{ji}^{+,s}}+\frac{\hat{A}_{ji}^{-,s}}{l_i-l_{ji}^{-,s}}\right)\label{Eq:partialf}\\
&\equiv \sum\limits_{s=\pm 1}\sum\limits_{i=1}^n\left(\,\int\limits_{-\infty}^{-1}+\int\limits_1^\infty\,\right)\mathrm{d}l_i\ n_\mathrm{F}\left(s M\frac{l_i^2+1}{l_i^2-1}\right)\hat{\mathcal{I}}_s(l_i,Q_{ji})\ ,
\end{align}
where obviously
\begin{equation}
\hat{A}_{ji}^{\pm,s}\equiv \lim_{l\rightarrow l_{ji}^{\pm,s}} (l-l_{ji}^{\pm,s})\hat{\mathcal{I}}_s(l,Q_{ji})\ .
\end{equation}
And explicitly,
\begin{eqnarray}
\hat{A}_{ji}^{\pm,s}&=&\frac{1}{2\pi}\frac{1}{(-s)}(M-\tilde{\Ksl}^s_{(ij\pm)})\notag\\
&&\times\left[\prod\limits_{\tiny\begin{array}{c}k=1\\k\neq i,j\end{array}}^n\frac{M-(\tilde{\Ksl}^s_{(ij\pm)}+\Qsl_{ki})}{\frac{1}{(l_{ji}^{\pm,s})^2-1} (\mathrm{i}2sMq_{ki,2}+Q_{ki}^2)(l_{ji}^{\pm,s}-l_{ki}^{+,s})(l_{ji}^{\pm,s}-l_{ki}^{-,s})}\right]\notag\\
&&\times\frac{M-(\tilde{\Ksl}^s_{(ij\pm)}+\Qsl_{ji})}{(\mathrm{i}2sMq_{ji,2}+Q_{ji}^2)(l_{ji}^{\pm,s}-l_{ji}^{\mp,s})}\ ,
\end{eqnarray}
where $\tilde{K}^s_{(ij\pm)}$ means making the substitution $l_i\mapsto l_{ji}^{\pm,s}$ in $\tilde{K}^s_{(i)}$. From Eq. (\ref{Eq:zeros}) it is evident that the momentum $\tilde{K}^s_{(ij\pm)}+Q_{ji}$ is also on-shell. And using the relation (\ref{Eq:spinorprod}), rotated to Euclidean space, we can rewrite this expression as
\begin{eqnarray}
\hat{A}_{ji}^{\pm,s}&=&\frac{1}{\mp 4s\pi Q_{ji}^2\sqrt{1+4M^2/Q_{ji}^2}}\, u(\tilde{K}^s_{(ij\pm)})\bar{u}(\tilde{K}^s_{(ij\pm)})u(\tilde{K}^s_{(ij\pm)}+Q_{ji})\bar{u}(\tilde{K}^s_{(ij\pm)}+Q_{ji})\notag\\
&&\times\left[\prod\limits_{\tiny\begin{array}{c}k=1\\k\neq i,j\end{array}}^n\frac{M-(\tilde{\Ksl}^s _{(ij\pm)}+\Qsl_{ki})}{(\tilde{\Ksl}^s_{(ij\pm)}+Q_{ki})^2+M^2}\right]\notag\\
&\equiv&\frac{\hat{T}_{ji}^{\pm,s}}{\mp 4s\pi Q_{ji}^2\sqrt{1+4M^2/Q_{ji}^2}}\ .\label{Eq:tree}
\end{eqnarray}
Also note that the momenta $\tilde{K}^s_{(ij\pm)}+Q_{ki}$ with $k\neq i,j$ instead, are not on-shell. Now, since $\tilde{K}_{(ij\pm)}$ and $\tilde{K}_{(ij\pm)}+Q_{ji}$ are both on-shell, this implies\footnote{Because then $(2\tilde{K}^s_{(ij\pm)}+Q_{ji})\cdot Q_{ji}=0$, and $(-2\tilde{K}^s_{(ji-)}+Q_{ji})\cdot Q_{ji}=0$. Thus, $2\tilde{K}^s_{(ij+)}+Q_{ji}=\alpha(2\tilde{K}^s_{(ij-)}+Q_{ji})$ (in two dimensions there are only two linearly-independent vectors), so $\alpha=-1$ (after looking at (\ref{Eq:tildeK}) and (\ref{Eq:zeros})), and therefore $\tilde{K}^s_{(ij-)}=-\tilde{K}^s_{(ij+)}-Q_{ji}$. On the other hand, $2\tilde{K}^s_{(ij+)}+Q_{ji}=\alpha(-2\tilde{K}^s_{(ji-)}+Q_{ji})$, so $\alpha=-1$, and $\tilde{K}^s_{(ji-)}=\tilde{K}^s_{(ij+)}+Q_{ji}$.}
\begin{equation}\label{Eq:condK}
\tilde{K}^s_{(ij+)}+Q_{ji}=\tilde{K}^s_{(ji-)}\ .
\end{equation}
Then, from (\ref{Eq:tree}) and (\ref{Eq:condK}) we get
\begin{equation}
\hat{T}_{ji}^{+,s}=\hat{T}_{ij}^{-,s}\equiv\hat{T}^s_{ji}\ .
\end{equation}
Furthermore, from the property $\tilde{K}_{(ij\pm)}\equiv\tilde{K}^{s=1}_{(ij\pm)}=-\tilde{K}^{s=-1}_{ji\mp}$, in a way analogous to the previous case it is not difficult to obtain
\begin{align}
\hat{T}_{ji}^{s=1}=\hat{T}_{ij}^{s=-1}\equiv\hat{T}_{ji}\ .
\end{align}

Then, working out the integral in (\ref{Eq:partialf}) and using the previous symmetry relations, after a tedious (although straightforward) simplification, one arrives to the final result
\begin{align}\label{Eq:fermioncut}
L_n=\sum\limits_{\tiny\begin{array}{c}i,j=1\\i\neq j\end{array}}^n\frac{\hat{T}_{ji}}{4\pi}\left\{\frac{1}{\beta(Q_{ji}^2)Q_{ji}^2}\ln\left[\frac{\beta(Q_{ji}^2)+1}{\beta(Q_{ji}^2)-1}\right] -2\int\limits_{-\infty}^\infty\mathrm{d}k\,\frac{n_\mathrm{F}(E_k)}{E_k}\,\frac{Q_{ji}^2+2kq_{ji,1}}{(Q_{ji}^2+2kq_{ji,1})^2+4E_k^2q_{ji,2}^2}\right\}\ ,
\end{align}
with $\beta(Q_{ji}^2)\equiv\sqrt{1+4M^2/Q_{ji}^2}$. The result of \cite{Ber77} is then obtained by particularizing for $T=0$ and performing the rotation to Minkowski space.

\begin{figure}[h!]
\begin{center}
\includegraphics[width=11cm]{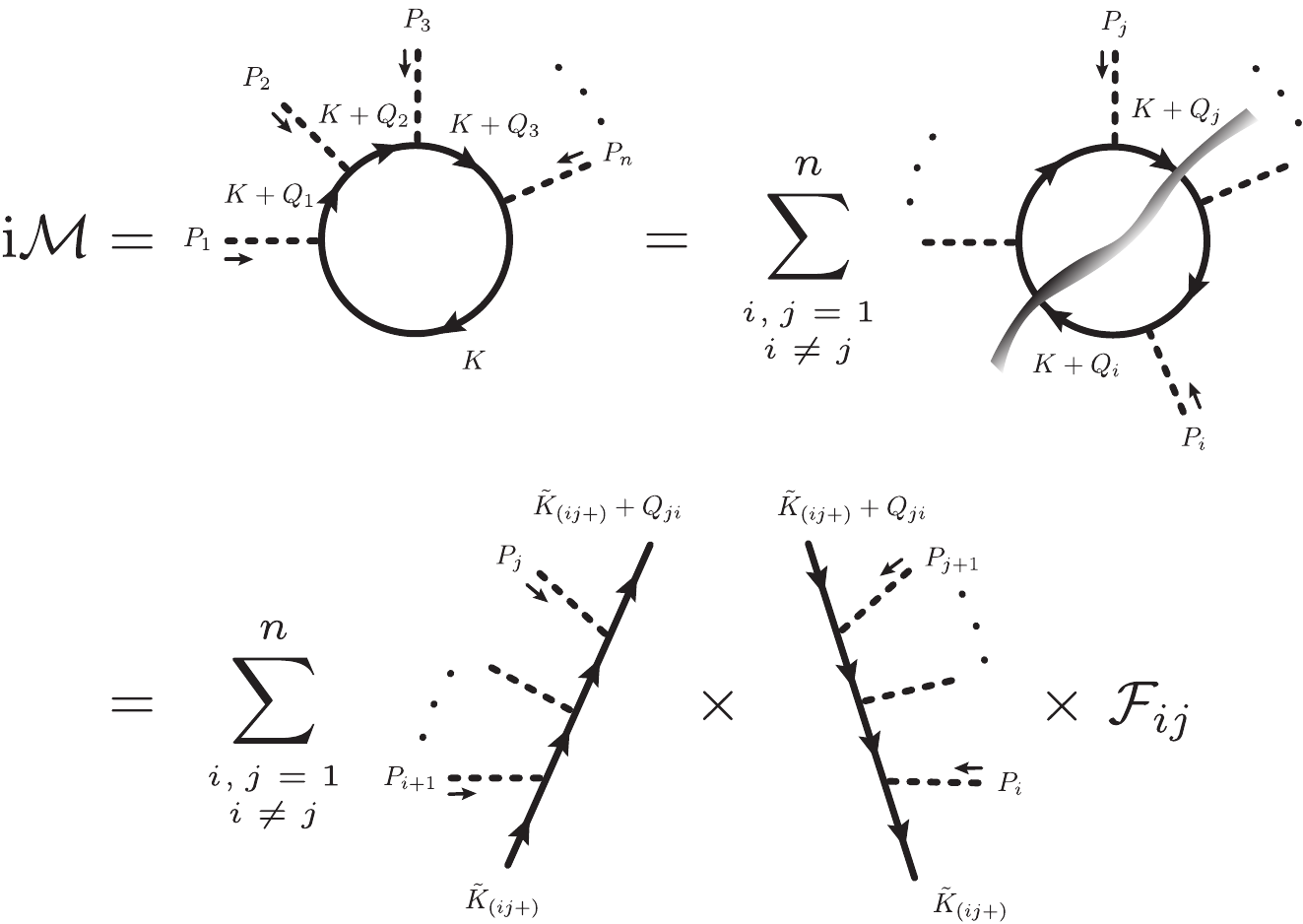}
\caption{Diagrammatic representation of the factorization of fermion loops in $1+1$ dimensions into two tree graphs, as implied by the result (\ref{Eq:fermioncut}). The momenta $\tilde{K}_{(ij+)}$ and $\tilde{K}_{(ij+)}+Q_{ji}$ are on-shell.}\label{Fig:fermioncut}
\end{center}
\end{figure}

We can represent the result of (\ref{Eq:fermioncut}) diagrammatically as in Fig. \ref{Fig:fermioncut}, with\footnote{Note that the contraction of Dirac indices due to the trace in the original loop diagram of Fig. (\ref{Fig:fermion-loop}) is apparently gone, but it is actually there because the external momenta of the trees are equal at both sides, and therefore
\begin{align}
&[\bar{u}(q)\tilde{A}u(p)][\bar{u}(p)\tilde{B}u(q)]=[\bar{u}_a(q)A_{ab}u_b(p)][\bar{u}_c(p)B_{cd}u_d(q)]\notag\\
&=u_d(q)\bar{u}_a(q)A_{ab}u_b(p)\bar{u}_c(p)B_{cd}=\Tr\{u(q)\bar{u}(q)\tilde{A}u(p)\bar{u}(p)\tilde{B}\}\ ,\label{Eq:tracespinors}
\end{align}
for any momenta $p,q$ and any matrices $\tilde{A},\tilde{B}$.}
\begin{align}\label{Eq:defF}
\mathcal{F}_{ij}\equiv\frac{1}{4\pi\beta(Q_{ji}^2)Q_{ji}^2}\ln\left[\frac{\beta(Q_{ji}^2)+1}{\beta(Q_{ji}^2)-1}\right] -\int\limits_{-\infty}^\infty\mathrm{d}k\,\frac{n_\mathrm{F}(E_k)}{2\pi E_k}\,\frac{Q_{ji}^2+2kq_{ji,1}}{(Q_{ji}^2+2kq_{ji,1})^2+4E_k^2q_{ji,2}^2}\ .
\end{align}

%
\section{Scattering amplitudes}\label{App:amplitudes}
As we saw in Section \ref{Sec:kinetictheory}, in $1+1$ dimensions only $3\rightarrow 3$ and $2\leftrightarrow 4$ scattering amplitudes contribute to the bulk viscosity at leading order in the $1/N$ expansion. Among them, in the large-$N$ limit, processes involving three different flavors dominate because of an overall factor $N^3$ obtained after summing over all flavor types.

By following the discussion in Section \ref{Sec:integrability}, we realize that the diagram of Fig. \ref{Fig:leadinginel}(g), after being cut in all possible ways, cancels the contribution from the diagrams \ref{Fig:leadinginel}(a--f), except for a term proportional to the integrability-breaking parameter, i.e., $m/M(T)$. And this also happens in the case of diagrams contributing to the $3\rightarrow 3$ amplitude, which are of the same order as the $2\leftrightarrow 4$ ones and have the same topology.

\begin{figure}[h!]
\begin{center}
\includegraphics[width=12cm]{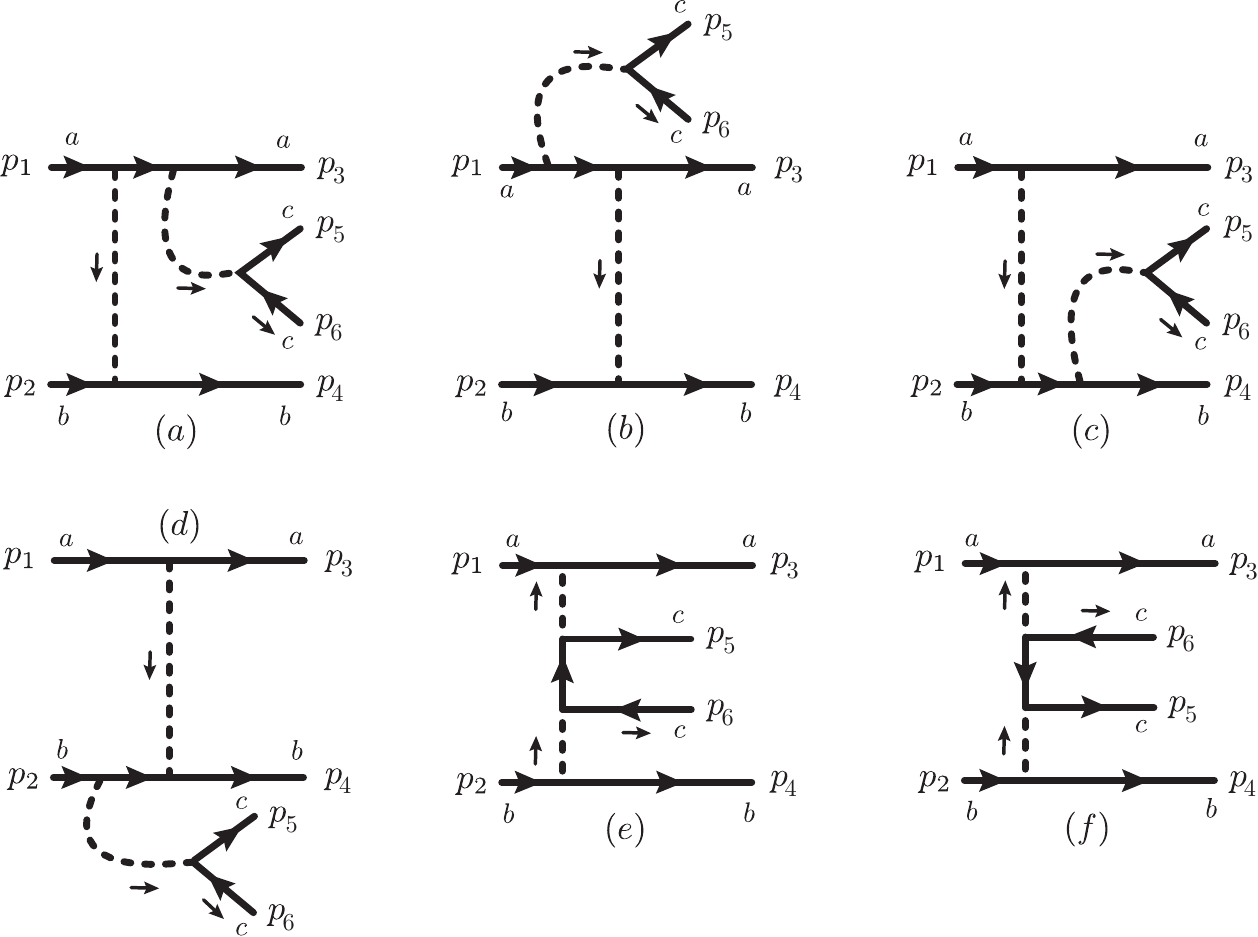}
\caption{Diagrams contributing to the inelastic process $\mathrm{f\,f\rightarrow\bar{f}\,f\,f\,f}$ in the massive Gross-Neveu model. Here $a$, $b$, and $c$ denote arbitrary flavors. The diagrams corresponding to $\mathrm{\bar{f}\,f\rightarrow\bar{f}\,\bar{f}\,f\,f}$ are obtained inverting the fermionic flow (but not the momentum) in the line which joins $p_1$ and $p_3$.}\label{Fig:fermion-fermion}
\end{center}
\end{figure}

\begin{figure}[h!]
\begin{center}
\includegraphics[width=12cm]{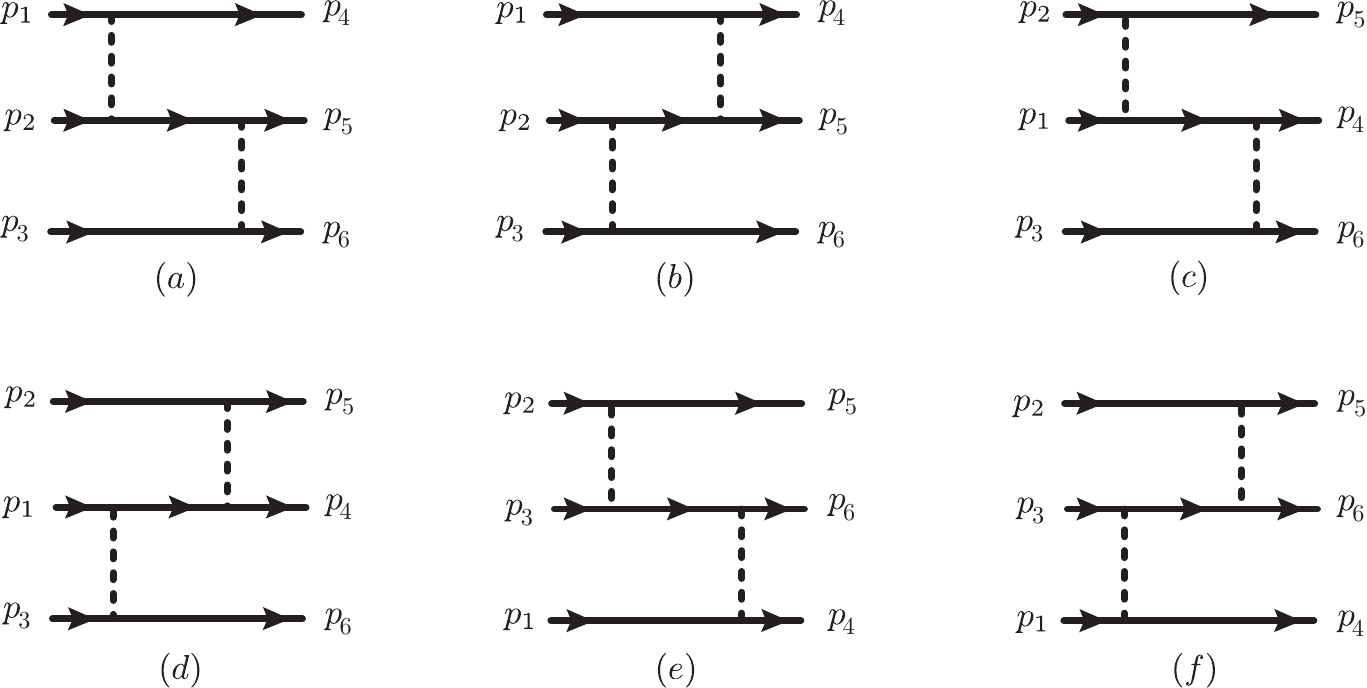}
\caption{Diagrams contributing to the elastic process $\mathrm{f\,f\,f\rightarrow f\,f\,f}$ in the massive Gross-Neveu model. Here $a$, $b$, and $c$ denote arbitrary flavors. The diagrams corresponding to $\mathrm{\bar{f}\,f\,f\rightarrow\bar{f}\,f\,f}$ are obtained inverting the fermionic flow (but not the momentum) in the line which joins $p_1$ and $p_4$.}\label{Fig:fermion-fermion-fermion}
\end{center}
\end{figure}

Consequently, the fermion-fermion inelastic and three-fermion elastic scattering amplitudes are given by the sum of the diagrams of Fig. \ref{Fig:fermion-fermion} and \ref{Fig:fermion-fermion-fermion} respectively times the integrability-breaking factor with a minus sign:
\begin{align}
& \mathrm{i}\mathcal{M}_\mathrm{f\,f\rightarrow\bar{f}\,f\,f\,f}=-\mathrm{i}\frac{m}{M(T)}\left[\mathcal{M}_\mathrm{f\, f\rightarrow\bar{f}\,f\,f\,f}^{(a)}+\mathcal{M}_\mathrm{f\, f\rightarrow\bar{f}\,f\,f\,f}^{(b)}+\mathcal{M}_\mathrm{f\, f\rightarrow\bar{f}\,f\,f\,f}^{(c)}+\mathcal{M}_\mathrm{f\, f\rightarrow\bar{f}\,f\,f\,f}^{(d)}+\mathcal{M}_\mathrm{f\, f\rightarrow\bar{f}\,f\,f\,f}^{(f)}\right]\ ,\\
& \mathrm{i}\mathcal{M}_\mathrm{f\,f\,f\rightarrow f\,f\,f}=-\mathrm{i}\frac{m}{M(T)}\left[\mathcal{M}_\mathrm{f\, f\,f\rightarrow f\,f\,f}^{(a)}+\mathcal{M}_\mathrm{f\, f\, f\rightarrow f\,f\,f}^{(b)}+\mathcal{M}_\mathrm{f\, f\,f\rightarrow f\,f\,f}^{(c)}+\mathcal{M}_\mathrm{f\, f\,f\rightarrow f\,f\,f}^{(d)}+\mathcal{M}_\mathrm{f\, f\,f\rightarrow f\,f\,f}^{(f)}\right]\ ,
\end{align}
and similarly for the processes involving more anti-fermions\footnote{By symmetry under charge conjugation, $|\mathcal{M}_\mathrm{f\,f\rightarrow\bar{f}\,f\,f\,f}|^2=|\mathcal{M}_\mathrm{\bar{f}\,\bar{f}\rightarrow\bar{f}\,\bar{f}\,\bar{f}\,f}|^2$, $|\mathcal{M}_\mathrm{f\,f\,f\rightarrow f\,f\,f}|^2=|\mathcal{M}_\mathrm{\bar{f}\,\bar{f}\,\bar{f}\rightarrow \bar{f}\,\bar{f}\,\bar{f}}|^2$, and $|\mathcal{M}_\mathrm{\bar{f}\,f\,f\rightarrow \bar{f}\,f\,f}|^2=|\mathcal{M}_\mathrm{f\,\bar{f}\,\bar{f}\rightarrow f\,\bar{f}\,\bar{f}}|^2$.}.


For instance, the explicit contribution from the diagram \ref{Fig:fermion-fermion}(a) is
\begin{align}
\mathrm{i}\mathcal{M}_\mathrm{f\, f\rightarrow\bar{f}\,f\,f\,f}^{(a)}=\bar{u}(p_3)S^\mathrm{ret}_\mathrm{F}(p_3+p_5+p_6)u(p_1)\bar{u}(p_4)u(p_2)\bar{u}(p_5)v(p_6) D_\sigma^\mathrm{ret}(p_4-p_2)D_\sigma^\mathrm{ret}(p_5+p_6)\ ,\label{Eq:Ma}
\end{align}
and analogously for the rest of diagrams. We already see from (\ref{Eq:Ma}), since $D_\sigma^\mathrm{ret}=\mathcal{O}(1/N)$, that the amplitude squared is $|\mathcal{M}_\mathrm{f\, f\rightarrow\bar{f}\,f\,f\,f}|^2=\mathcal{O}(1/N^2)$ (and the same for the other amplitudes).

It is not difficult to realize that the momenta in the argument of the retarded fermion propagators cannot be on-shell, being each $p_i$ on-shell, therefore we can make the substitution $S_\mathrm{F}^\mathrm{ret}\mapsto S_\mathrm{F}(p_0,p_1)=1/(\psl-M)$ in our calculations without worrying about singularities. After summing all the amplitudes and squaring them, we can simplify the expression using (\ref{Eq:spinorprod}), the property $D_{\sigma,\mathrm{ret}}^*(p)=D_{\sigma,\mathrm{ret}}(-p)$, and the relations for traces in Dirac space
\begin{eqnarray}
\Tr\left\{(\psl+M)(\qsl+M)\right\}&=&2(M^2+p\cdot q)\ ,\\
\Tr\left\{(\psl+M)(\ksl+M)(\qsl+M)\right\}&=&2M[M^2+p\cdot(k+q)+k\cdot q]\ ,\\
\Tr\left\{(\psl+M)(\ksl+M)(\qsl+M)(\lsl+M)\right\}&=&2[M^2(M^2+l\cdot(p+k+q)+p\cdot(k+q)+k\cdot q)\notag\\
&&+(l\cdot p)(k\cdot q)-(l\cdot k)(p\cdot q)+(l\cdot q)(p\cdot k)]\ .
\end{eqnarray}
Since the final expressions for the squared amplitudes after simplification are still very long and do not explicitly provide further significant information, I avoid to write them down here.


\end{document}